\documentclass[
aps,superscriptaddress,citesort,floatfix,nofootinbib,nolongbibliography,
]{revtex4-2}

\usepackage[utf8]{inputenc}
\usepackage[T1]{fontenc}

\usepackage{graphicx}
\usepackage{bm}

\usepackage{amsmath, amssymb}

\usepackage{booktabs,threeparttable}
\usepackage{tabularx}

\usepackage[top=2cm,bottom=2cm,left=1.75cm,right=1.75cm]{geometry}

\usepackage{hyperref}
\usepackage{xcolor}
\definecolor{tab_blue}{HTML}{1F77B4}
\hypersetup{%
  breaklinks=true,
  colorlinks=true,
  linkcolor=tab_blue,
  filecolor=tab_blue,
  urlcolor=tab_blue,
  citecolor=tab_blue,
}

\def\be{\begin{equation}}
\def\ee{\end{equation}}
\def\ba{\begin{eqnarray}}
\def\ea{\end{eqnarray}}
\def\epa{\bm{e}_\parallel}
\def\kpn{k_{\perp}}
\def\ppn{p_{\perp}}
\def\qpn{q_{\perp}}
\def\kpa{k_{\parallel}}
\def\kk{\bm{k}}
\def\pp{\bm{p}}
\def\qq{\bm{q}}
\def\kknb{\bm{k}_\perp}
\def\ppnb{\bm{p}_\perp}
\def\qqnb{\bm{q}_\perp}

\def\vv{\bm{v}}
\def\ww{\bm{w}}
\def\bnabla{\bm{\nabla}}

\newcommand{\beq}{\begin{equation}}
\newcommand{\eeq}{\end{equation}}
\newcommand{\bea}{\begin{eqnarray}}
\newcommand{\eea}{\end{eqnarray}}
\newcommand{\nn}{\nonumber}

\begin{document}

\title{Wave turbulence theory of odd fluids and solids: kinetic equations and solutions}
\author{Xander M. de Wit}
\thanks{Both authors contributed equally to this work.}
\affiliation{Department of Applied Physics and Science Education, Eindhoven University of Technology, 5600 MB Eindhoven, Netherlands}
\author{L\'eo Touzo}
\thanks{Both authors contributed equally to this work.}
\affiliation{James Franck Institute, The University of Chicago, Chicago, IL 60637, USA}
\author{Sébastien Galtier}
\affiliation{Université Paris-Saclay, Laboratoire de Physique des Plasmas, École polytechnique, 91128 Palaiseau, France}
\author{Michel Fruchart}
\affiliation{Gulliver, ESPCI Paris, Université PSL, CNRS, 75005 Paris, France}
\author{Federico Toschi}
\affiliation{Department of Applied Physics and Science Education, Eindhoven University of Technology, 5600 MB Eindhoven, Netherlands}
\affiliation{CNR-IAC, I-00185 Rome, Italy}
\author{Vincenzo Vitelli}
\email{vitelli@uchicago.edu}
\affiliation{James Franck Institute, The University of Chicago, Chicago, IL 60637, USA}
\affiliation{Leinweber Institute for Theoretical Physics, The University of Chicago, Chicago, IL 60637, USA}

\date{\today}

\begin{abstract}
The theory of wave turbulence describes the properties of physical systems composed of a set of weak-amplitude random waves interacting nonlinearly. Here, we study odd wave turbulence, which arises in chiral media subjected to non-reciprocal stresses, notably odd viscosity  and odd elasticity. In both cases, we consider simple models for which we can derive and solve analytically the kinetic equations describing the long-term statistical behavior of spectral quantities such as energy or wave action. For odd viscosity, we consider a three-dimensional model that exhibits wave turbulence involving three-wave interactions, which gives rise to a direct energy cascade characterized by an anisotropic Kolmogorov-Zakharov (KZ) spectrum. For odd elasticity, we consider a quasi-one-dimensional overdamped model that exhibits much slower dynamics involving six-wave interactions. In that case, the KZ spectrum corresponding to a forward cascade of a conserved quantity we call odd energy, is nonlocal and therefore does not constitute a physical solution. However, the other KZ solution, which describes an inverse cascade of wave action, is only marginally non-local and is therefore valid up to a logarithmic correction. 
These two analytical theories provide a rigorous interpretation of direct numerical simulations, where the KZ spectrum is observed both in the case of odd viscosity (forward cascade) and of odd elasticity (inverse cascade).
\end{abstract}

\maketitle

\section{Introduction}

Wave turbulence, the statistical description of interacting (nonlinear) wave systems, 
finds applications in fields such as gravity waves, plasma waves, Bose-Einstein condensation, elastic waves on a plate or gravitational waves \cite{Zakharov1992,Nazarenko11,Newell2011,GaltierCUP2023}. An important feature of such systems is the occurrence of energy cascades through scales, when a large scale separation exists between driving and dissipation.
In this paper, we study wave turbulence in two different types of parity-violating nonequilibrium media involving chiral forces: fluids with odd viscosity and solids with odd elasticity. In both of these cases, energy is constantly injected and dissipated at the microscopic scale, which strongly affects the physical properties of the system at the macroscopic level and makes it out of equilibrium even in the absence of a turbulent cascade.

The first model we study is a three-dimensional chiral fluid with odd viscosity. 
Odd viscosity describes situations where a fluid exhibits transverse responses to velocity gradients~\cite{oddreview}. Such an effect has been measured experimentally in a variety of systems ranging from two dimensional magnetized graphene~\cite{Berdyugin2019} to spinning colloids~\cite{Soni2019} and three-dimensional magnetized polyatomic gases \cite{Beenakker1970}, while strong but indirect experimental evidence also exists in 3D magnetized plasma \cite{Stacey2006,Bae2013,Kono2015} (under the name of {\it gyroviscosity}) . 

The second model we study is a quasi-one-dimensional elongated solid with odd elasticity. Odd elasticity consists in a transverse response to the deformation of a solid \cite{Scheibner2020b,oddreview}. While usual elastic solids allow for the propagation of waves in the presence of inertia, waves can propagate through an odd elastic solid even in the overdamped regime, which provides a distinct ground for the study of wave turbulence. 
This behavior has been reported in systems ranging from robotic solids \cite{Chen2021,Veenstra2025} to artificial or living colloidal systems \cite{Bililign2021,Tan2022,Chao2026} and to biological systems \cite{Shankar2024}.

We show based on numerical simulations and phenomenological arguments that a weak wave turbulent cascade can indeed be observed in both models.
In the case of odd fluids~(Ref.~\cite{nhwt}), energy conservation at the mesoscopic scale is preserved by odd viscosity, and we observe a modified energy cascade.
Contrary to odd viscosity, odd elasticity requires the input and dissipation of energy, which is therefore not conserved even at the macroscopic level.
Instead, we find that an inverse turbulent cascade may still occur for an emergent conserved quantity called wave action.

In order to describe analytically the behavior of these systems, we turn to wave turbulence theory. 
The wave kinetic equation (WKE) is a central tool for describing 
interacting wave systems, both in equilibrium and non-equilibrium states. 
The main assumption for the rigorous derivation of the WKE is that the waves have a weak amplitude and that, therefore, the nonlinear interaction is weak. Under this assumption, we can, for example, use a multiple time scale approach to derive the WKE \cite{Benney1966,Newell1968,Galtier2024}. 
This equation describes the time evolution of spectral quantities corresponding to inviscid invariants of the system (such as energy, wave action or momentum). The significance of wave turbulence lies in the fact that, starting from the WKE, one can derive non-trivial stationary spectra corresponding to finite flux solutions \cite{Zakharov1967}. These exact solutions are called Kolmogorov-Zakharov (KZ) spectra and describe either a direct cascade or an inverse cascade. The wave turbulence regime and its solutions have been observed in numerous experiments and direct numerical simulations \cite{Yarom2014,Cosmo2017,during2017wave,Meyrand2018,Hassaini2019,LeReun2020,Monsalve2020,Galtier2021,Ricard2021,Falcon2022,Griffin2022,Rodda2022,David2024,Kochurin2024,Zhao2024}. 

\begin{figure}
    \centering
    \includegraphics[width=0.45\linewidth,trim={10cm 1.8cm 10cm 1.1cm},clip]{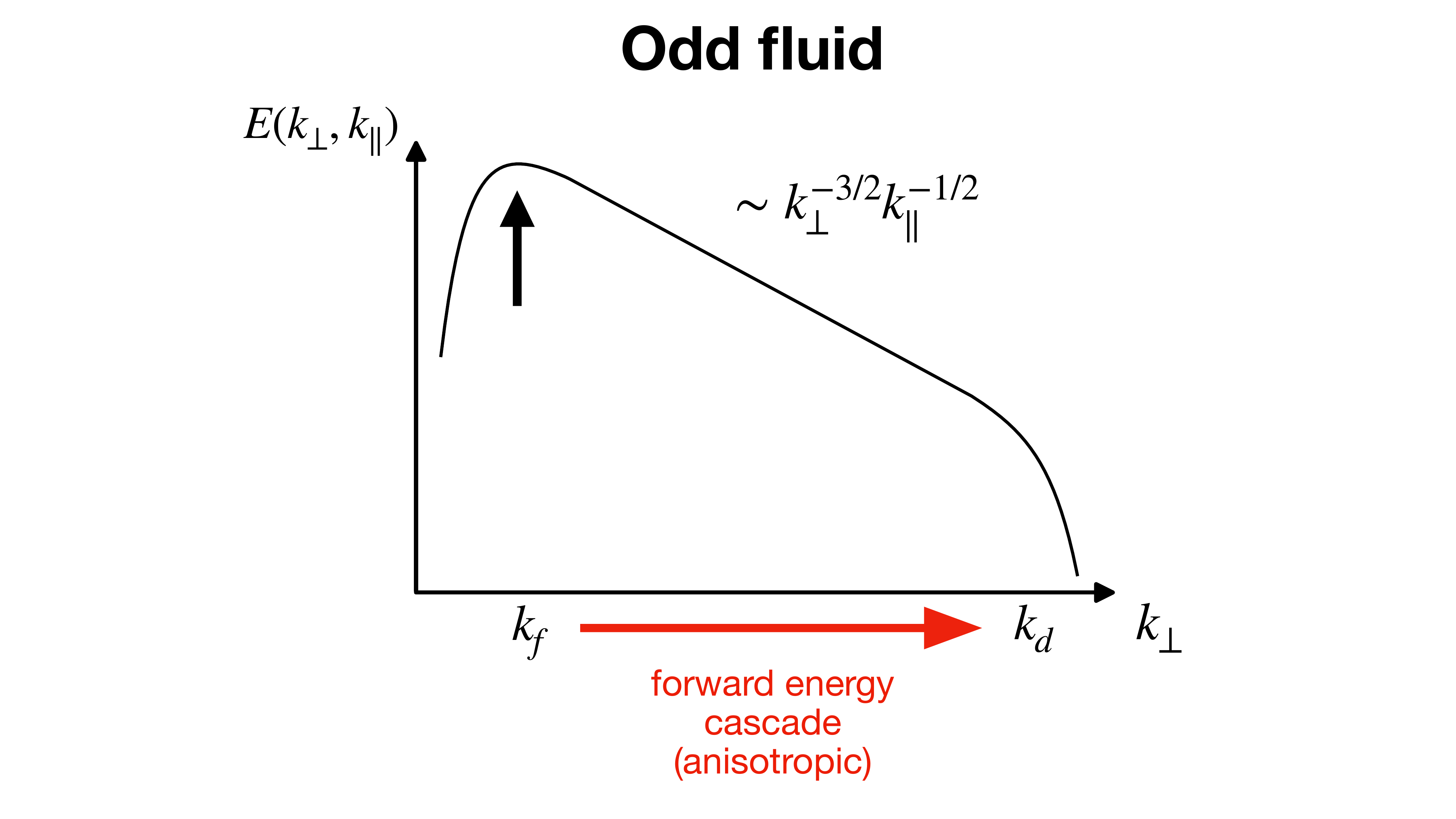}
    \hspace{0.8cm}
    \includegraphics[width=0.45\linewidth,trim={10cm 1.8cm 10cm 1.1cm},clip]{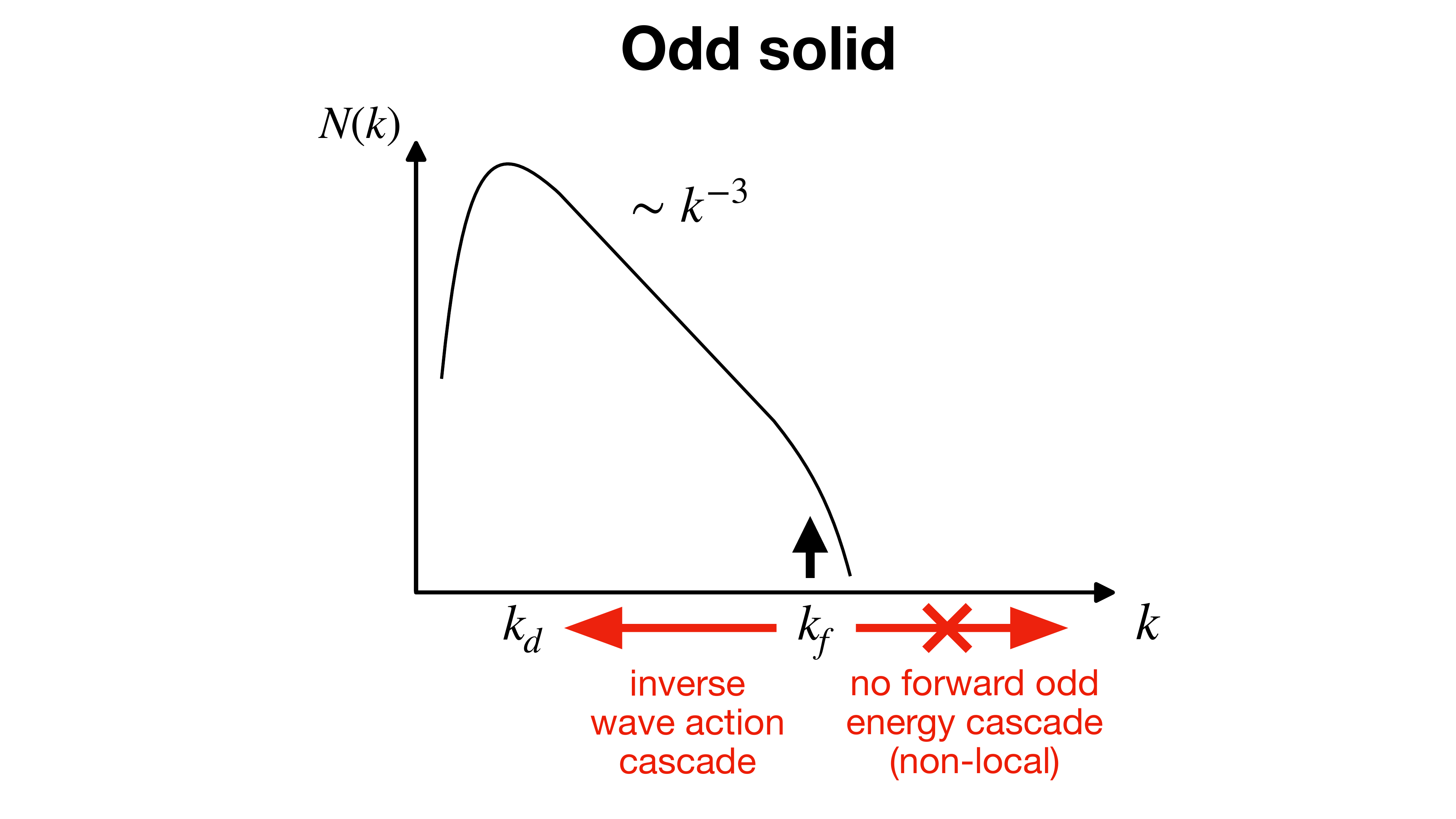}
    \caption{
    \textbf{Summary of the analytical results of weak wave turbulence theory.}
    Left: Sketch of the anisotropic energy spectrum for the odd viscous fluid (Eq.~\eqref{eq:NS}). When energy is injected at large scales, a forward cascade develops associated to a KZ spectrum $E(k_\perp,k_|)\propto k_\perp^{-3/2} k_\|^{-1/2}$. Right: Sketch of the wave action spectrum for the odd elastic solid (Eq.~\eqref{eom_odd_solid_true}). When wave action is injected at small or intermediate scales, an inverse cascade develops associated to a KZ spectrum $N(k)\propto k^{-3}$. A forward cascade of odd energy is in principle possible, but it is not present due to the associated KZ solution being non-local.}
    \label{fig:summary}
\end{figure}

The purpose of the present article is to provide a detailed explanation of the derivation of the wave kinetic equation for the two models sketched above, as well as its solutions and their properties. 
In Section~\ref{sec:odd_fluid}, we focus on the weak wave turbulence regime of the odd viscous fluid model. This allows us in particular to obtain from a rigorous analytical calculation the expression of the energy spectrum observed in the numerical simulations shown in Ref.~\cite{nhwt}. In Section~\ref{sec:odd_solid}, we consider the odd elastic solid. In the weak regime, our analytical calculations suggest the absence of a forward cascade, the corresponding solution of the WKE being non-local. An inverse cascade, corresponding to a flux of wave action from small to large scales is however possible, and we were indeed able to observe it in our numerical simulations. Beyond the weak regime, strong turbulence may also arise in this system, with two distinct regimes where the spectrum exhibits different exponents.
Our results are summarized in Fig.~\ref{fig:summary}.

\section{Wave turbulence in an odd viscous fluid} \label{sec:odd_fluid}

\subsection{Model and phenomenology}

In this first part of the paper, we consider an incompressible chiral fluid with cylindrical symmetry along a preferred direction $\bm{e}_\parallel$ \cite{Khain2022}, with constituent particles spinning about this axis. The velocity field $\bm{v}$ obeys the Navier–Stokes equations with odd viscosity, 
\begin{gather}
\label{eq:NS}
     \partial_t\bm{v} + (\bm{v}\cdot\bm{\nabla}) \bm{v} = - \bm{\nabla} p + 
     \begin{pmatrix}
         \nu & \nu_\textrm{odd} & 0 \\
         -\nu_\textrm{odd} & \nu & 0 \\
         0 & 0 & \nu
     \end{pmatrix}
     \Delta \bm{v}
+ \bm{f}
\qquad
\text{with}
\qquad
\bm{\nabla}\cdot\bm{v}=0 \;,
\end{gather}
where we have set the density to 1. Here $p$ is the reduced pressure, $\bm{f}$ the driving force, $\nu$ the shear viscosity, and $\nu_\textrm{odd}$ the odd viscosity coefficient.
For $\nu_\textrm{odd} \neq 0$, the matrix in Eq.~\eqref{eq:NS} is non-symmetric, reflecting the non-reciprocal nature of odd viscosity. The wavevector is decomposed as $\bm{k} = (k_\perp \cos\theta, k_\perp \sin\theta, k_\parallel)$ in the basis $(\bm{e}_\perp^1,\bm{e}_\perp^2,\bm{e}_\parallel)$ in which the matrix in Eq.~\eqref{eq:NS} is expressed. Odd viscosity acts as a scale-dependent Coriolis-like term, $-\nu_{\textrm{odd}} k^2 \bm{e}_\parallel \times \bm{v}$, whose strength increases with wavenumber $k = |\bm{k}|$. At the linear level, this permits so-called odd waves with dispersion relation $\omega_{\bm{k}} = \pm \nu_\textrm{odd} k_\parallel k$ (see below)~\cite{Avron1998,dewit2024,chen2024}.

If odd viscosity is small, these odd waves are slow and do not influence the dynamics, yielding the regular eddy-turbulence that is known from the Navier-Stokes equations without odd viscosity. However, as odd viscosity increases and the timescales of the odd waves enter in the inertial range of turbulence, the odd waves can drastically alter the dynamics~\cite{dewit2024,chen2024,nhwt}. These linear odd waves interact through the non-linear term on the left-hand side of Eq.~\eqref{eq:NS}. These two regimes, respectively called the {\it strong} turbulence and {\it weak} wave turbulence regime, can be characterized through the ratio $\chi \equiv \tau_{\textrm{lin}} / \tau_{\textrm{NL}}$ between the linear time scale $\tau_{\textrm{lin}}\equiv1/\omega_{\bm{k}}$ and the non-linear time scale (or eddy timescale) $\tau_\textrm{NL} \equiv 1/(k \sqrt{E(k_\perp,k_\parallel) k_\perp k_\parallel})$. The strong regime is typically realized when $\chi>\mathcal{O}(1)$. This occurs when energy is injected at large scales \mbox{$k_f<k_\textrm{odd}\equiv \varepsilon^{1/4}\nu_\textrm{odd}^{-3/4}$}, where $\varepsilon$ is the turbulent energy injection rate \cite{dewit2024}. In this case, Kolmogorov-like turbulence is observed in the range $k<k_\textrm{odd}$ where waves are slow ($\chi > \mathcal{O}(1)$). Below the crossover scale, for $k>k_\textrm{odd}$, the eddy timescale and wave timescale become of the same order ($\chi\sim\mathcal{O}(1)$), which corresponds to a state called critical balance \cite{Higdon1984,GS95,Nazarenko2011,Oughton2020,Zhou2021,Alexakis2018}. In this range, it has been shown that energy accumulation gives rise to pattern formation, with a spectral scaling $E(k)\sim k^{-1}$ \cite{dewit2024, chen2024}.

If instead we inject energy at smaller scales $k_f > k_\textrm{odd}$, eddy turbulence still occurs inside the 2D manifold corresponding to $k_\|=0$, leading to an inverse cascade of energy in the range $k<k_f$, as explained in Ref.~\cite{dewit2024}. Nevertheless, this upscale energy transfer only involves a portion of the kinetic energy. A significant fraction of the energy flux is instead cascaded forward into the range $k>k_f$ (similar to the situation in inertial wave turbulence \cite{Shaltiel2024}). The dynamics of this spectral region remain largely unexplored. As we will verify a posteriori below, this situation where $k>k_f > k_\textrm{odd}$ is the range where $\chi \ll \mathcal{O}(1)$ for all modes. This means that the odd linear waves dominate the dynamics, and they only weakly interact through the non-linear term. Their dynamics can hence be understood from the framework of weak wave turbulence (in the 3D manifold, with $k_\parallel>0$). Below we will employ weak wave turbulence theory to derive exact solutions for the energy spectrum and energy cascade of this system.

\subsection{Conservation of energy}

Let us show that the odd viscosity conserves the kinetic energy of the system, 
\be
\mathcal{E} = \frac{1}{2} \int \bm{dr} \, \bm{v}^2 \;. 
\ee
Using \eqref{eq:NS}, we can write
\be
\partial_t \mathcal{E} = \int \bm{dr} \, \bm{v} \cdot \partial_t \bm{v} = \int \bm{dr} \, \bm{v} \cdot [-\bm{v}\cdot \bm{\nabla} \bm{v} - \bm{\nabla} p + \begin{pmatrix}
         \nu & \nu_\textrm{odd} & 0 \\
         -\nu_\textrm{odd} & \nu & 0 \\
         0 & 0 & \nu
     \end{pmatrix}
     \Delta \bm{v} + \bm{f} ] \;.
\ee
Both the advection term and the pressure term vanish due to the incompressibility, as
\be
-\int \bm{dr} \, \bm{v} \cdot (\bm{v}\cdot \bm{\nabla} \bm{v}) = - \frac{1}{2} \int \bm{dr} \, \bm{v}\cdot \bm{\nabla} (\bm{v}^2) = \frac{1}{2} \int \bm{dr} \, \bm{v}^2 \, \bm{\nabla} \cdot \bm{v} = 0
\ee
and
\be
-\int \bm{dr} \,  \bm{v}\cdot \bm{\nabla} p = \int \bm{dr} \, p \,\bm{\nabla} \cdot \bm{v} = 0
\ee
(we assume that the velocity is zero at the boundaries such that the boundary terms vanish). Similarly, it is easy to show that the contribution due to the odd viscosity vanishes. The viscosity term reads
\bea
\int \bm{dr} \, [ \nu \, \bm{v} \cdot \Delta \bm{v} + \nu_\textrm{odd} u_x \Delta u_y - \nu_\textrm{odd} u_y \Delta u_x] &=& \int \bm{dr} \, [ \nu \, \bm{v} \cdot \Delta \bm{v} - \nu_\textrm{odd} (\bm{\nabla} u_x) \cdot (\bm{\nabla} u_y) + \nu_\textrm{odd} (\bm{\nabla} u_y) \cdot (\bm{\nabla} u_x) ] \\
&=& \int \bm{dr} \, \nu \, \bm{v} \cdot \Delta \bm{v} \;. \nn
\eea
Thus, we see that the addition of odd viscosity does not affect energy conservation. We thus finally obtain
\be
\partial_t \mathcal{E} = \int \bm{dr} \, \bm{v} \cdot [ \nu \Delta \bm{v} + \bm{f} ] \;.
\ee
In this paper, we will consider a situation where the driving force only acts on large scales, and where the shear viscosity is small, such that the dissipation is restricted to very small scales. There is thus a wide range of scales, called the inertial range, for which the energy is conserved. In this range, a cascade of energy may develop, with energy flowing from the large scales to the small scales.

\subsection{Theoretical results}

In this section, we derive the WKE of a fluid with odd viscosity. We also present the assumptions, definitions and properties of this type of turbulence. Exact solutions are also discussed. 

\subsubsection{Wave amplitude equation}
\label{section2}

The Navier-Stokes equations with odd viscosity $\nu_\textrm{odd}$ can be written in vector form
\be 
\frac{\partial \bm{v}}{\partial t} + (\bm{v} \cdot \bnabla)  \bm{v} = - \bnabla p + \nu \Delta  \bm{v} + \nu_\textrm{odd} \epa \times \Delta \bm{v} \;, \label{NS1}
\ee
It is convenient to rewrite this system for the vorticity field $\ww$; we obtain
\be
\frac{\partial \ww}{\partial t} +  \nu_\textrm{odd} (\epa \cdot \bnabla) \Delta \bm{v} = (\ww \cdot \bnabla)  \bm{v} - (\bm{v} \cdot \bnabla) \ww + \nu \Delta  \ww \;. 
\label{NSrot}
\ee
Hereafter, we will neglect the term proportional to the viscosity $\nu$. 
\\

\noindent\textbf{Canonical variables.}
We introduce the toroidal ($\psi$) and poloidal ($\phi$) scalar fields in the following manner
\be
\bm{v} = \bnabla \times (\psi \epa) + \bnabla \times (\bnabla \times (\phi \epa) ) \;,
\ee
whose Fourier transform reads (with $k=\vert \kk \vert$)
\be
\hat \vv_{\kk} = i \hat \psi_{\kk} \kk \times \epa -  \hat \phi_{\kk} \kk \times (\kk \times \epa ) 
= i \hat \psi_{\kk} \kk \times \epa + \hat \phi_{\kk} ( k^2 \epa - \kpa \kk ) \;,
\ee
and from which we obtain the vorticity vector
\be
\hat \ww_{\kk} = \hat \psi_{\kk} ( k^2 \epa- \kpa \kk) + i k^2 \hat \phi_{\kk}  \kk \times \epa \;. 
\ee
In Fourier space, the linear contribution of equation (\ref{NSrot}) leads, after projection, to 
\begin{subequations}
\ba
\frac{\partial \hat \psi_{\kk}}{\partial t} &=& i \nu_\textrm{odd} \kpa k^2 \hat \phi_{\kk} \;, \\
\frac{\partial \hat \phi_{\kk}}{\partial t} &=& i \nu_\textrm{odd} \kpa \hat \psi_{\kk} \;.
\ea
\end{subequations}
The linear solutions of this equation are helical odd waves with angular frequency
\be
\omega^2_{\kk} = \nu_\textrm{odd}^2 \kpa^2 k^2 .
\ee
Using this property, we introduce the canonical variables 
\be \label{vcanon}
A^{s}_{\kk} \equiv A^{s} (\kk) = k \hat \psi_{\kk} - s k^2 \hat \phi_{\kk} \;,
\ee
with $s=\pm$ the directional polarity. With such a choice of canonical variables, we have
\be
\vert A^+_{\kk} \vert^2 + \vert A^-_{\kk} \vert^2 = 2 \vert \hat \vv_{\kk} \vert^2 
\ee
and at the linear level
\be
\frac{\partial A^s_{\kk}}{\partial t} + i s \omega_{\kk} A^s_{\kk} = 0 \quad , \quad \omega_{\kk} = \nu_\textrm{odd} \kpa k \;.
\ee
Note that throughout this section, $k_\perp$ denotes the norm of $\kk_\perp=\kk-(\kk \cdot{\bm e}_\|){\bm e}_\|$ and is always positive, while $k_\|=\kk \cdot{\bm e}_\|$ denotes a component of $\kk$ which can be negative. 
\\

\noindent\textbf{Resonance condition.}
The resonance condition for three-wave interactions between waves with frequencies $s\omega_{\kk}$, $s_p\omega_{\pp}$, $s_q\omega_{\qq}$ and wavevectors $\kk$, $\pp$, $\qq$, respectively, reads \citep{GaltierCUP2023}
\begin{subequations} \label{defreson}
\ba
s \omega_{\kk} + s_p \omega_{\pp} + s_q \omega_{\qq} &=& 0 \;, \\
\kk + \pp + \qq &=& 0 \;.
\ea
\end{subequations}
In the case of odd waves, these relations are equivalent to the conditions
\be \label{reson}
\frac{s_q q - s_p p}{k_\parallel} = \frac{sk - s_q q}{p_\parallel} = \frac{s_p p - sk}{q_\parallel} . 
\ee
When the system is initially excited at large scale in a narrow isotropic domain in Fourier space (a situation often considered in numerical simulations) the dynamics is initially dominated by local interactions such that $k \simeq p \simeq q$. As the locality of the interactions is a property of turbulence that is generally verified, we extend its use beyond the initial instant and get
\be
\frac{s_q -s_p}{\kpa} \simeq \frac{s - s_q}{p_\parallel} \simeq \frac{s_p - s}{q_\parallel} . 
\ee
From this expression, we can show that the associated cascade is necessarily anisotropic. Indeed, if $\kpa$ is non-zero, the left-hand term only gives a non-negligible contribution when $s_{p}=-s_{q}$. As a consequence, either the middle or the right-hand term has a vanishing numerator (to leading order), which implies that the corresponding denominator must also cancel (to leading order) to satisfy the equality.
For instance, if $s=s_{p}$ then $q_\parallel \simeq 0$. This condition means that the transfer in the parallel direction is negligible. Indeed, the integration of the wave amplitude equation in the parallel direction (see below) is then reduced to a few modes (since $p_\parallel \simeq \kpa$) which strongly limits the transfer between parallel modes. The cascade in the parallel direction is therefore possible but relatively weak compared to that in the perpendicular direction. In the following, we will take advantage of this property and consider the anisotropic limit $\kpn \gg \kpa$ to simplify the derivation. 
We can still use the locality condition when turbulence is anisotropic with $k \sim k_\perp$. We then obtain $k_\perp \sim p_\perp \sim q_\perp$, whereas the parallel wavenumbers are limited to a narrow domain.
\\

\noindent\textbf{Wave amplitude equation.}
In the derivation of the wave amplitude equation, we will consider a continuous medium which can lead to mathematical difficulties connected with  infinite dimensional phase spaces. For this reason, it is preferable to assume a variable spatially periodic over a box of finite size $L$. However, in the derivation of the kinetic equation, the limit $L \to +\infty$ is finally taken (before the long time limit). As both approaches lead to the same kinetic equation, for simplicity, we anticipate this result and follow the original approach of \citet{Benney1966}. 
Note that the anisotropic limit ($\kpn \gg \kpa$) will also be taken before the (asymptotic) long time limit. 

The first non-linear term of equation (\ref{NSrot}) writes
\ba
\widehat{(\ww \cdot \bnabla) \vv}_{\kk} &=& i \int (\hat \ww_{\pp} \cdot \qq) \hat \vv_{\qq} \delta_{\kk}^{\pp\qq} d\pp d\qq \nonumber \\
&=& i \int \left[ i \hat \phi_{\pp} \hat \phi_{\qq} p^2 \left( \qq \cdot (\pp \times \epa)\right) (q^2 \epa -q_\parallel \qq) 
- \hat \phi_{\pp} \hat \psi_{\qq} p^2 \left( \qq \cdot (\pp \times \epa)\right) (\qq \times \epa) \right. \nonumber \\
&&\left. \mbox{} - \hat \psi_{\pp} \hat \phi_{\qq} \left( p_\parallel \pp \cdot \qq - p^2 q_\parallel \right) (q^2 \epa -q_\parallel \qq) 
- i \hat \psi_{\pp} \hat \psi_{\qq} \left( p_\parallel \pp \cdot \qq - p^2 q_\parallel \right) (\qq \times \epa)  \right] \nonumber \\
&& \mbox{}\times \delta_{\kk}^{\pp\qq} d\pp d\qq \;,
\ea
with $\delta_{\kk}^{\pp\qq} \equiv \delta(\kk-\pp-\qq)$. In the anisotropic limit ($\kpn \gg \kpa$), a first simplification arises
\ba
\widehat{(\ww \cdot \bnabla) \vv}_{\kk} &=& i \int \left[ i \hat \phi_{\pp} \hat \phi_{\qq} \ppn^2 \qpn^2 \left( \epa \cdot (\qqnb \times \ppnb)\right) \epa
- \hat \phi_{\pp} \hat \psi_{\qq} \ppn^2 \left( \epa \cdot (\qqnb \times \ppnb)\right) (\qqnb \times \epa) \right. \nonumber \\
&&\left. \mbox{} - \hat \psi_{\pp} \hat \phi_{\qq} \qpn^2 \left( p_\parallel \ppnb \cdot \qqnb - \ppn^2 q_\parallel \right) \epa 
- i \hat \psi_{\pp} \hat \psi_{\qq} \left( p_\parallel \ppnb \cdot \qqnb - \ppn^2 q_\parallel \right) (\qqnb \times \epa)  \right] \nonumber \\
&& \mbox{} \times \delta_{\kk}^{\pp\qq} d\pp d\qq \;.
\ea

The second non-linear term of equation (\ref{NSrot}) reads
\ba
\widehat{(\vv \cdot \bnabla) \ww}_{\kk} &=& i \int (\hat \vv_{\pp} \cdot \qq) \hat \ww_{\qq} \delta_{\kk}^{\pp\qq} d\pp d\qq \nonumber \\
&=& i \int \left[ i \hat \phi_{\pp} \hat \phi_{\qq} \left( p^2 q_\parallel -p_\parallel \pp \cdot \qq \right) q^2 (\qq \times \epa) 
+ \hat \phi_{\pp} \hat \psi_{\qq} \left( p^2 q_\parallel - p_\parallel \pp \cdot \qq \right) (q^2 \epa -q_\parallel \qq) \right. \nonumber \\
&&\left. \mbox{} - \hat \psi_{\pp} \hat \phi_{\qq} \left( \qq \cdot (\pp \times \epa)  \right) q^2 (\qq \times \epa)
+ i \hat \psi_{\pp} \hat \psi_{\qq} \left( \qq \cdot (\pp \times \epa) \right) (q^2 \epa - q_\parallel \qq) \right] \nonumber \\
&& \mbox{} \times \delta_{\kk}^{\pp\qq} d\pp d\qq \;,
\ea
which simplifies in the anisotropic limit to
\ba
\widehat{(\vv \cdot \bnabla) \ww}_{\kk} &=& i \int \qpn^2 \left[ i \hat \phi_{\pp} \hat \phi_{\qq} \left( \ppn^2 q_\parallel -p_\parallel \ppnb \cdot \qqnb \right) (\qqnb \times \epa) 
+ \hat \phi_{\pp} \hat \psi_{\qq} \left( \ppn^2 q_\parallel - p_\parallel \ppnb \cdot \qqnb \right) \epa \right. \nonumber \\
&&\left. \mbox{} - \hat \psi_{\pp} \hat \phi_{\qq} \left( \epa \cdot (\qqnb \times \ppnb) \right) (\qqnb \times \epa)
+ i \hat \psi_{\pp} \hat \psi_{\qq} \left( \epa \cdot (\qqnb \times \ppnb) \right) \epa \right] \nonumber \\
&& \mbox{} \times \delta_{\kk}^{\pp\qq} d\pp d\qq \;.
\ea
The addition of these two non-linear contributions leads to the simplified expression
\ba
\widehat{NL}(\kk) &=& \widehat{(\ww \cdot \bnabla) \vv}_{\kk} - \widehat{(\vv \cdot \bnabla) \ww}_{\kk} \nonumber \\
&=&\int \hat \phi_{\pp} \hat \phi_{\qq} \ppn^2\qpn^2 \left( \epa \cdot (\ppnb \times \qqnb)\right) \epa \delta_{\kk}^{\pp\qq} d\pp d\qq \nonumber \\
&& \mbox{} +i \int \hat \phi_{\pp} \hat \psi_{\qq} \ppn^2 \left( \epa \cdot (\ppnb \times \qqnb)\right) (\qqnb \times \epa) \delta_{\kk}^{\pp\qq} d\pp d\qq  \nonumber \\
&& \mbox{} -i \int \hat \psi_{\pp} \hat \phi_{\qq} \qpn^2 \left( \epa \cdot (\ppnb \times \qqnb) \right) (\qqnb \times \epa) \delta_{\kk}^{\pp\qq} d\pp d\qq  \nonumber \\
&& \mbox{} - \int \hat \psi_{\pp} \hat \psi_{\qq} \qpn^2 \left( \epa \cdot (\ppnb \times \qqnb) \right) \epa \delta_{\kk}^{\pp\qq} d\pp d\qq \;. 
\ea
The introduction of the canonical variables 
\begin{subequations}
\ba
\hat \psi_{\kk} &=& \frac{1}{2 \kpn} \sum_s A_{\kk}^s \;, \\
\hat \phi_{\kk} &=& - \frac{1}{2 \kpn^2} \sum_s sA_{\kk}^s \;,
\ea
\end{subequations}
gives
\ba
\widehat{NL}(\kk) &=& \frac{1}{4} \sum_{s_p s_q} \int s_p s_q A_{\pp}^{s_p} A_{\qq}^{s_q} \left( \epa \cdot (\ppnb \times \qqnb)\right) \epa \delta_{\kk}^{\pp\qq} d\pp d\qq \nonumber \\
&&\mbox{} -\frac{i}{4} \sum_{s_p s_q} \int A_{\pp}^{s_p} A_{\qq}^{s_q} \frac{s_p}{\qpn} \left( \epa \cdot (\ppnb \times \qqnb)\right) (\qqnb \times \epa) \delta_{\kk}^{\pp\qq} d\pp d\qq  \nonumber \\
&&\mbox{}  + \frac{i}{4} \sum_{s_p s_q} \int A_{\pp}^{s_p} A_{\qq}^{s_q} \frac{s_q}{\ppn} \left( \epa \cdot (\ppnb \times \qqnb) \right) (\qqnb \times \epa) \delta_{\kk}^{\pp\qq} d\pp d\qq  \nonumber \\
&&\mbox{}  - \frac{1}{4} \sum_{s_p s_q} \int A_{\pp}^{s_p} A_{\qq}^{s_q} \frac{\qpn}{\ppn} \left( \epa \cdot (\ppnb \times \qqnb) \right) \epa \delta_{\kk}^{\pp\qq} d\pp d\qq \;. 
\ea
The dummy variables $\pp$, $\qq$ and $s_p$, $s_q$, can be exchanged to symmetrize the equation; we find
\ba
\widehat{NL}(\kk) 
&=& \frac{1}{8} \sum_{s_p s_q} \int A_{\pp}^{s_p} A_{\qq}^{s_q} \frac{\epa \cdot (\ppnb \times \qqnb)}{\ppn \qpn}
(\ppn^2-\qpn^2) \epa \delta_{\kk}^{\pp\qq} d\pp d\qq \\
&&\mbox{} +\frac{i}{8} \sum_{s_p s_q} \int A_{\pp}^{s_p} A_{\qq}^{s_q} \frac{\epa \cdot (\ppnb \times \qqnb)}{\ppn \qpn}
\left( s_q \qpn - s_p \ppn \right)(\kknb \times \epa) \delta_{\kk}^{\pp\qq} d\pp d\qq \;. \nonumber
\ea
Coming back to the wave amplitude equation, we can write in the anisotropic limit
\ba
\left( \frac{\partial \hat \psi_{\kk}}{\partial t} - i \nu_\textrm{odd} k_\parallel \kpn^2 \hat \phi_{\kk} \right) \kpn^2 \epa
+ \left( i \frac{\partial \hat \phi_{\kk}}{\partial t} + \nu_\textrm{odd} \kpa \hat \psi_{\kk} \right) \kpn^2 \kknb \times \epa = \widehat{NL}(\kk) \;,
\ea
therefore, after projection and use of the dispersion relation, we obtain
\begin{subequations}
\ba
\frac{\partial \hat \psi_{\kk}}{\partial t} - i \omega_{\kk} \kpn \hat \phi_{\kk} &=&
\sum_{s_p s_q} \int A_{\pp}^{s_p} A_{\qq}^{s_q} \frac{\epa \cdot (\ppnb \times \qqnb)}{8 \kpn^2 \ppn \qpn} (\ppn^2-\qpn^2) \delta_{\kk}^{\pp\qq} d\pp d\qq \;, \quad \\
\frac{\partial \hat \phi_{\kk}}{\partial t} - i \omega_{\kk} \frac{\hat \psi_{\kk}}{\kpn} &=& 
\sum_{s_p s_q} \int A_{\pp}^{s_p} A_{\qq}^{s_q} \frac{\epa \cdot (\ppnb \times \qqnb)}{8 \kpn^2 \ppn \qpn} \left( s_q \qpn - s_p \ppn \right) \delta_{\kk}^{\pp\qq} d\pp d\qq \;. \quad
\ea
\end{subequations}
With the introduction of the canonical variables (\ref{vcanon}), the weighted addition of the previous expressions gives
\ba
\frac{\partial A^s_{\kk}}{\partial t} + i s \omega_{\kk} A^s_{\kk} &=& \sum_{s_p s_q} \int \frac{\epa \cdot (\ppnb \times \qqnb)}{8 \kpn \ppn \qpn} 
\left( \ppn^2-\qpn^2 - s \kpn \left( s_q \qpn - s_p \ppn \right) \right) \nonumber \\
&& \mbox{} \times A_{\pp}^{s_p} A_{\qq}^{s_q} \delta_{\kk}^{\pp\qq} d\pp d\qq \;.
\ea
Remarking that
\be
\ppn^2-\qpn^2 = (s_p \ppn - s_q \qpn)(s_p\ppn + s_q \qpn) \;,
\ee
we can rearrange the expression in the following manner
\ba
\frac{\partial A^s_{\kk}}{\partial t} + i s \omega_{\kk} A^s_{\kk} &=&
\sum_{s_p s_q} \int \frac{\epa \cdot (\ppnb \times \qqnb)}{8 \kpn \ppn \qpn} ( s_p \ppn - s_q \qpn) ( s \kpn + s_p \ppn + s_q \qpn) \nonumber \\
&& \mbox{} \times A_{\pp}^{s_p} A_{\qq}^{s_q} \delta_{\kk}^{\pp\qq} d\pp d\qq \;.
\ea
We introduce the interaction representation for waves of weak amplitude ($0 < \epsilon \ll 1$)
\be \label{A_to_a}
A_{\kk}^s = \epsilon a_{\kk}^s e^{-is\omega_{\kk} t} \;,
\ee
and get eventually the wave amplitude equation after a few last manipulations
\be \label{WE}
\frac{\partial a^s_{\kk}}{\partial t} = \epsilon \sum_{s_p s_q} \int L_{\kk\pp\qq}^{ss_ps_q} a_{\pp}^{s_p} a_{\qq}^{s_q} e^{i \Omega_{\kk,\pp\qq} t} \delta_{\kk}^{\pp\qq} d\pp d\qq \;,
\ee
with $\Omega_{\kk,\pp\qq} \equiv s \omega_{\kk} - s_p \omega_{\pp} - s_q \omega_{\qq}$ and 
\be \label{coeff}
L_{\kk\pp\qq}^{ss_ps_q} \equiv \frac{\epa \cdot (\ppnb \times \qqnb)}{8 \kpn \ppn \qpn} ( s_p \ppn - s_q \qpn ) 
( s \kpn + s_p \ppn + s_q \qpn) \;.
\ee
Expression (\ref{coeff}) satisfies the following properties on the resonant manifold defined by \eqref{defreson}\footnote{Note that if the resonant manifold is defined instead by $s \omega_{\kk} - s_p \omega_{\pp} - s_q \omega_{\qq} = 0$ and $\kk - \pp - \qq = 0$ as in \eqref{WE}, the relations \eqref{reson} and \eqref{Lprop} have to be adapted through a change $(k_\|,p_\|,q_\|)\to (k_\|,-p_\|,-q_\|)$.} (relation (\ref{reson}) is used)
\begin{subequations} \label{Lprop}
\ba
L_{{\bm 0}\pp\qq}^{ss_ps_q} &=& 0 , \\
L_{\kk\qq\pp}^{ss_qs_p} &=& L_{\kk\pp\qq}^{ss_ps_q} , \\
L_{\pp\kk\qq}^{s_pss_q} &=& \frac{p_\parallel}{k_\parallel} L_{\kk\pp\qq}^{ss_ps_q} , \\
L_{\kk\pp\qq}^{-s-s_p-s_q} &=& L_{\kk\pp\qq}^{ss_ps_q} , \\
L_{-\kk-\pp-\qq}^{ss_ps_q} &=& L_{\kk\pp\qq}^{ss_ps_q} . 
\ea
\end{subequations}
The wave amplitude equation (\ref{WE}) governs the slow evolution of odd waves of weak amplitude in the anisotropic limit. It is a quadratic non-linear equation which corresponds to the interactions between waves propagating along $\pp$ and $\qq$, in the positive ($s_{p},s_{q}>0$) or negative  ($s_{p},s_{q}<0$) direction. The symmetries listed above can be used to simplify the derivation of the kinetic equation \citep{Galtier2023b}.
The wave amplitude equation tells us that the non-linear coupling between the states associated with the wavevectors $\ppnb$ and $\qqnb$ vanishes when these wavevectors are collinear. Moreover, we note that the non-linear coupling disappears when the wavenumbers $\ppn$ and $\qpn$ are equal if their associated directional polarities, $s_p$ and $s_q$, are also equal. 
These are general properties for helical waves \citep{Kraichnan1973,Waleffe1992,Turner2000,Galtier2003,Galtier2006,Galtier2014}.

\subsubsection{Wave kinetic equation}
\label{sectionKE}

The derivation of the WKE for odd wave turbulence is classical. The method based on a multiple time scale was recently reviewed for inertial wave turbulence, a similar problem where waves are helical and turbulence anisotropic \citep{Galtier2023b}. Note that this technique was first proposed by \citet{Benney1966} for three-wave interactions and then extended to four-wave interactions \citep{Newell1968} with an application to surface gravity waves.

Assuming a statistically homogeneous turbulence, we introduce the energy density spectrum $e_{\kk}^{s} \delta (\kk-\kk') \equiv \langle a^{s}_{\kk} a^{s*}_{\kk'} \rangle$, with $e_{\kk}^{s} \equiv e^{s}(\kk)$. Using the result of \citep{Galtier2023b}, the multiple time scale method leads to
\ba \label{KEIW_step1}
\frac{\partial e_{\kk}^{s}}{\partial t} &=& 4\pi\epsilon^2 \sum_{s_{p} s_{q}} \int |L_{\kk\pp\qq}^{ss_ps_q}|^2 e_{\pp}^{s_p} e_{\qq}^{s_q} \delta(\Omega_{\kk,\pp\qq}) \delta_{\kk}^{\pp\qq} d\pp d\qq  \\
&&\mbox{} + 8\pi\epsilon^2 \sum_{s_{p} s_{q}} \int L_{\kk\pp\qq}^{ss_ps_q} L_{\pp-\qq\kk}^{s_ps_qs} e_{\kk}^{s} e_{\qq}^{s_q} \delta(\Omega_{\kk,\pp\qq}) \delta_{\kk}^{\pp\qq} d\pp d\qq \;.\nonumber
\ea
Using the properties of the coefficient $L_{\kk\pp\qq}^{ss_ps_q}$ given in \eqref{Lprop}, as well as the symmetry $e^s(\bm{k})=e^{s}(-\bm{k})$, and finally using the symmetry between $\bm{p}$ and $\bm{q}$, we obtain
the following kinetic equation
\ba \label{KEIW}
\frac{\partial e_{\kk}^{s}}{\partial t} &=& \frac{\pi \epsilon^2 k_\parallel}{16} \sum_{s_{p} s_{q}} \int  
\left( \frac{\sin \theta_k}{\kpn} \right)^2 \left(\frac{s_p \ppn - s_q \qpn}{k_\parallel} \right)^2 ( s \kpn + s_p \ppn + s_q \qpn)^2 \\
&&\mbox{} \times \left[ k_\parallel e_{\pp}^{s_p} e_{\qq}^{s_q} + p_\parallel e_{\kk}^{s} e_{\qq}^{s_q} + q_\parallel e_{\kk}^{s} e_{\pp}^{s_p} \right] 
\delta (\Omega_{\kk\pp\qq}) \delta_{\kk\pp\qq} d\pp d\qq \;,  \nonumber
\ea
where $\Omega_{\kk\pp\qq}=s \omega_{\kk} + s_p \omega_{\pp} + s_q \omega_{\qq}$, and $\theta_k$ is the opposite angle to $\kknb$ in the triangle $\kknb+\ppnb+\qqnb = {\bf 0}$. The expression of $\sin \theta_k$ can be obtained via the law of cosines,
\be
k_\perp^2 = p_\perp^2+q_\perp^2 - 2p_\perp q_\perp \cos \theta_k \;,
\ee
which gives
\be \label{sinethetak}
\sin \theta_k = \frac{1}{2 p_\perp q_\perp}\sqrt{2p_\perp^2 q_\perp^2 + 2k_\perp^2 p_\perp^2 + 2k_\perp^2 q_\perp^2 - k_\perp^4 - p_\perp^4 - q_\perp^4} \;.
\ee
Expression (\ref{KEIW}) is the kinetic equation for odd wave turbulence in the anisotropic limit ($k_\parallel \ll k_\perp$). Note that the kinetic equation for odd wave turbulence does not describe the slow mode ($\kpa=0$), which involves strong turbulence.

\subsubsection{Detailed conservation laws}

The kinetic equation satisfies the conservation of energy and helicity. To prove this property, we introduce the energy spectrum
\be
E(\kk) = e^+(\kk)+ e^-(\kk) = \sum_s e^s(\kk)
\ee
and the helicity spectrum 
\be
H(\kk) = \kpn ( e^+(\kk)- e^-(\kk) )  = \kpn \sum_s s e^s(\kk) \;.
\ee
For the energy, we find
\ba \label{consE}
&&\frac{\partial \int E(\kk) d\kk}{\partial t} = \frac{\pi \epsilon^2}{16} \sum_{s s_{p} s_{q}} \int k_\parallel
\left( \frac{\sin \theta_k}{\kpn} \right)^2 \left(\frac{s_p \ppn - s_q \qpn}{k_\parallel} \right)^2 \\
&&\mbox{}  \times ( s \kpn + s_p \ppn + s_q \qpn)^2 \left[ k_\parallel e_{\pp}^{s_p} e_{\qq}^{s_q} + p_\parallel e_{\kk}^{s} e_{\qq}^{s_q} + q_\parallel e_{\kk}^{s} e_{\pp}^{s_p} \right] 
\delta (\Omega_{\kk\pp\qq}) \delta_{\kk\pp\qq} d\kk d\pp d\qq \;.  \nonumber
\ea
After a circular permutation we obtain
\ba \label{consE2}
&&\frac{\partial \int E(\kk) d\kk}{\partial t} = \frac{\pi \epsilon^2}{48} \sum_{s s_{p} s_{q}} \int (k_\parallel + p_\parallel + q_\parallel)
\left( \frac{\sin \theta_k}{\kpn} \right)^2 \left(\frac{s_p \ppn - s_q \qpn}{k_\parallel} \right)^2 \\
&&\mbox{} \times ( s \kpn + s_p \ppn + s_q \qpn)^2 \left[ k_\parallel e_{\pp}^{s_p} e_{\qq}^{s_q} + p_\parallel e_{\kk}^{s} e_{\qq}^{s_q} + q_\parallel e_{\kk}^{s} e_{\pp}^{s_p} \right] 
\delta (\Omega_{\kk\pp\qq}) \delta_{\kk\pp\qq} d\kk d\pp d\qq \;,  \nonumber
\ea
which is null on the resonant manifold. This proved the detailed (for each triad) energy conservation. 

Likewise, for the helicity we find
\ba \label{consH}
&&\frac{\partial \int H(\kk) d\kk}{\partial t} = \frac{\pi \epsilon^2}{16} \sum_{s s_{p} s_{q}} \int s k_\parallel \kpn 
\left( \frac{\sin \theta_k}{\kpn} \right)^2 \left(\frac{s_p \ppn - s_q \qpn}{k_\parallel} \right)^2 \\
&&\mbox{}  \times ( s \kpn + s_p \ppn + s_q \qpn)^2 \left[ k_\parallel e_{\pp}^{s_p} e_{\qq}^{s_q} + p_\parallel e_{\kk}^{s} e_{\qq}^{s_q} + q_\parallel e_{\kk}^{s} e_{\pp}^{s_p} \right] 
\delta (\Omega_{\kk\pp\qq}) \delta_{\kk\pp\qq} d\kk d\pp d\qq \;.  \nonumber
\ea
After a circular permutation we obtain
\ba \label{consH2}
&&\frac{\partial \int H(\kk) d\kk}{\partial t} = \frac{\pi \epsilon^2}{48} \sum_{s s_{p} s_{q}} \int (s \omega_k + s_p \omega_p + s_q \omega_q) 
\left( \frac{\sin \theta_k}{\kpn} \right)^2 \left(\frac{s_p \ppn - s_q \qpn}{k_\parallel} \right)^2 \\
&&\mbox{}  \times ( s \kpn + s_p \ppn + s_q \qpn)^2 \left[ k_\parallel e_{\pp}^{s_p} e_{\qq}^{s_q} + p_\parallel e_{\kk}^{s} e_{\qq}^{s_q} + q_\parallel e_{\kk}^{s} e_{\pp}^{s_p} \right] 
\delta (\Omega_{\kk\pp\qq}) \delta_{\kk\pp\qq} d\kk d\pp d\qq \;,  \nonumber
\ea
which is null on the resonant manifold. Therefore, energy and helicity are conserved by the kinetic equation. Note that usually the conservation of energy is obtained using the dispersion relation while the conservation of the second invariant is obtained using the wave vector relation. This is really an odd turbulence. 
\\

\subsubsection{Kolmogorov-Zakharov spectrum}

The derivation of the stationary solutions requires a long but classical calculation. First, we assume axisymmetry and introduce the reduced spectrum
\be
E^s_{\kk} \equiv E^s(\kpn,k_\parallel) = 2 \pi \kpn e^s(\kk) \;.
\ee
The kinetic equation \eqref{KEIW} now reads 
\ba \label{KEani_step}
\frac{\partial E_{\kk}^{s}}{\partial t} &=& \frac{\epsilon^2 k_\parallel}{32 \nu_\textrm{odd}} \sum_{s_{p} s_{q}} \int
 \frac{\sin^2 \theta_k}{\kpn} \frac{1}{\kpn \ppn \qpn}
\left(\frac{s_p \ppn - s_q \qpn}{k_\parallel} \right)^2 ( s \kpn + s_p \ppn + s_q \qpn)^2 \\
&&\mbox{} \times \left[ \omega_{\kk} E_{\pp}^{s_p} E_{\qq}^{s_q} + \omega_{\pp} E_{\kk}^{s} E_{\qq}^{s_q} + \omega_{\qq} E_{\kk}^{s} E_{\pp}^{s_p} \right] 
\delta (\Omega_{\kk\pp\qq}) \delta_{\kk\pp\qq} d\pp d\qq \;.  \nonumber
\ea
Performing an angular integration in the plane containing $\bm{p}_\perp$ and $\bm{q}_\perp$, we obtain
\ba \label{KEani}
\frac{\partial E_{\kk}^{s}}{\partial t} &=& \frac{\epsilon^2 k_\parallel}{32 \nu_\textrm{odd}} \sum_{s_{p} s_{q}} \int_{\Delta_\perp}
\left( \frac{\sin \theta_k}{\kpn} \right) \frac{1}{\kpn \ppn \qpn}
\left(\frac{s_p \ppn - s_q \qpn}{k_\parallel} \right)^2 ( s \kpn + s_p \ppn + s_q \qpn)^2 \\
&&\mbox{} \times \left[ \omega_{\kk} E_{\pp}^{s_p} E_{\qq}^{s_q} + \omega_{\pp} E_{\kk}^{s} E_{\qq}^{s_q} + \omega_{\qq} E_{\kk}^{s} E_{\pp}^{s_p} \right] 
\delta (\Omega_{\kk\pp\qq}) \delta(k_\parallel+p_\parallel+q_\parallel) d\ppn d\qpn dp_\parallel dq_\parallel \;,  \nonumber
\ea
where the integration domain $\Delta_\perp$ is an infinitely long band defined by the equations
\ba
p_\perp + q_\perp &>& k_\perp \;, \\
|p_\perp-q_\perp| &<& k_\perp \nn
\ea
(while $p_\parallel$ and $q_\parallel$ are still integrated over $\mathbb{R}$).
To simplify our analysis, we shall consider the case of zero-helicity, $E_{\kk}^+=E_{\kk}^-$. Then, we obtain
\ba \label{KEani2}
\frac{\partial E_{\kk}}{\partial t} &=& \frac{\epsilon^2 k_\parallel}{128 \nu_\textrm{odd}} \sum_{s s_{p} s_{q}} \int_{\Delta_\perp}
\left( \frac{\sin \theta_k}{\kpn} \right) \frac{1}{\kpn \ppn \qpn}
\left(\frac{s_p \ppn - s_q \qpn}{k_\parallel} \right)^2 ( s \kpn + s_p \ppn + s_q \qpn)^2 \\
&&\mbox{} \times \left[ \omega_{\kk} E_{\pp} E_{\qq} + \omega_{\pp} E_{\kk} E_{\qq} + \omega_{\qq} E_{\kk} E_{\pp} \right] 
\delta (\Omega_{\kk\pp\qq}) \delta(k_\parallel+p_\parallel+q_\parallel) d\ppn d\qpn dp_\parallel dq_\parallel \;.  \nonumber
\ea
We introduce the adimensionalized wave numbers $\tilde p_i \equiv p_i/k_i$ and $\tilde q_i \equiv q_i/k_i$ with $i=\perp,\parallel$. 
Assuming a spectrum of the type $E_{\kk} = A \kpn^n |k_\parallel|^m$, we finally obtain
\ba \label{KEani3}
\frac{\partial E_{\kk}}{\partial t} &=& \frac{\epsilon^2 A^2}{128 \nu_\textrm{odd}} \kpn^{2n+2} |k_\parallel|^{2m} \sum_{s s_{p} s_{q}} \int_{\Delta_\perp} 
\frac{\sin \theta_k}{\tilde p_\perp \tilde q_\perp}
\left(s_p \tilde p_\perp - s_q \tilde q_\perp \right)^2 ( s + s_p \tilde p_\perp + s_q \tilde q_\perp)^2 \\
&&\mbox{} \times \left[ {\tilde p_\perp}^n {\tilde q_\perp}^n |{\tilde p_\parallel}|^m |{\tilde q_\parallel}|^m +
{\tilde p_\perp} {\tilde q_\perp}^n {\tilde p_\parallel} |{\tilde q_\parallel}|^m +
{\tilde p_\perp}^n {\tilde q_\perp} |{\tilde p_\parallel}|^m {\tilde q_\parallel} \right] \nonumber \\
&&\mbox{} \times 
\delta (s+s_p \tilde p_\perp \tilde p_\parallel + s_q \tilde q_\perp \tilde q_\parallel) 
\delta(1+\tilde p_\parallel+\tilde q_\parallel) d\tilde p_\perp d\tilde q_\perp d\tilde p_\parallel d\tilde q_\parallel \;.  \nonumber
\ea
We now apply the Kuznetsov-Zakharov (KZ) transformation. We first perform the conformal change of variables
\be
\tilde p_i \to \frac{1}{\tilde p_i} \quad , \quad \tilde q_i \to \frac{\tilde q_i}{\tilde p_i} \quad , \quad s\leftrightarrow s_p \;,
\ee
with Jacobian $1/(\tilde p_\perp \tilde p_\|)^3$ (with again $i=\perp,\parallel$). Using \eqref{sinethetak}, this implies in particular
\be
\sin \theta_k = \frac{1}{2 \tilde p_\perp \tilde q_\perp}\sqrt{2\tilde p_\perp^2 \tilde q_\perp^2 + 2\tilde p_\perp^2 + 2\tilde q_\perp^2 - 1 - \tilde p_\perp^4 - \tilde q_\perp^4} \to \tilde p_\perp \sin \theta_k \;.
\ee
Using also the relations \eqref{reson}, this leads to
\ba \label{KEani3p}
\frac{\partial E_{\kk}}{\partial t} &=& \frac{\epsilon^2 A^2}{128 \nu_\textrm{odd}} \kpn^{2n+2} |k_\parallel|^{2m} \sum_{s s_{p} s_{q}} \int_{\Delta_\perp} 
\frac{\sin \theta_k}{\tilde p_\perp \tilde q_\perp}
\left(s_p \tilde p_\perp - s_q \tilde q_\perp \right)^2 ( s + s_p \tilde p_\perp + s_q \tilde q_\perp)^2 \\
&&\mbox{} \times \tilde p_\perp^{-3-2n} |p_\parallel|^{-2m} {\rm sign} (\tilde p_\|) \left[ {\tilde p_\perp}^n {\tilde q_\perp}^n |{\tilde p_\parallel}|^m |{\tilde q_\parallel}|^m +
{\tilde p_\perp} {\tilde q_\perp}^n {\tilde p_\parallel} |{\tilde q_\parallel}|^m +
{\tilde p_\perp}^n {\tilde q_\perp} |{\tilde p_\parallel}|^m {\tilde q_\parallel} \right] \nonumber \\
&&\mbox{} \times 
\delta (s+s_p \tilde p_\perp \tilde p_\parallel + s_q \tilde q_\perp \tilde q_\parallel) 
\delta(1+\tilde p_\parallel+\tilde q_\parallel) d\tilde p_\perp d\tilde q_\perp d\tilde p_\parallel d\tilde q_\parallel \;.  \nonumber
\ea
Applying instead the change of variables
\be
\tilde p_i \to \frac{\tilde p_i}{\tilde q_i} \quad , \quad \tilde q_i \to \frac{1}{\tilde q_i} \quad , \quad s\leftrightarrow s_q
\ee
gives
\ba \label{KEani3q}
\frac{\partial E_{\kk}}{\partial t} &=& \frac{\epsilon^2 A^2}{128 \nu_\textrm{odd}} \kpn^{2n+2} |k_\parallel|^{2m} \sum_{s s_{p} s_{q}} \int_{\Delta_\perp} 
\frac{\sin \theta_k}{\tilde p_\perp \tilde q_\perp}
\left(s_p \tilde p_\perp - s_q \tilde q_\perp \right)^2 ( s + s_p \tilde p_\perp + s_q \tilde q_\perp)^2 \\
&&\mbox{} \times \tilde q_\perp^{-3-2n} |q_\parallel|^{-2m} {\rm sign}( \tilde q_\|) \left[ {\tilde p_\perp}^n {\tilde q_\perp}^n |{\tilde p_\parallel}|^m |{\tilde q_\parallel}|^m +
{\tilde p_\perp} {\tilde q_\perp}^n {\tilde p_\parallel} |{\tilde q_\parallel}|^m +
{\tilde p_\perp}^n {\tilde q_\perp} |{\tilde p_\parallel}|^m {\tilde q_\parallel} \right] \nonumber \\
&&\mbox{} \times 
\delta (s+s_p \tilde p_\perp \tilde p_\parallel + s_q \tilde q_\perp \tilde q_\parallel) 
\delta(1+\tilde p_\parallel+\tilde q_\parallel) d\tilde p_\perp d\tilde q_\perp d\tilde p_\parallel d\tilde q_\parallel \;.  \nonumber
\ea
Summing together the expressions \eqref{KEani3}, \eqref{KEani3p} and \eqref{KEani3q}, we finally obtain
\ba \label{KEani4}
\frac{\partial E_{\kk}}{\partial t} &=& \frac{\epsilon^2 A^2}{384 \nu_\textrm{odd}} \kpn^{2n+2} |k_\parallel|^{2m} \sum_{s s_{p} s_{q}} \int_{\Delta_\perp}
\sin \theta_k \left(s_p \tilde p_\perp - s_q \tilde q_\perp \right)^2 ( s + s_p \tilde p_\perp + s_q \tilde q_\perp)^2 \, {\tilde p_\perp}^{n-1} {\tilde q_\perp}^{n-1} |{\tilde p_\parallel}|^m |{\tilde q_\parallel}|^m \\
&&\mbox{} \times
\left[ 1 + {\tilde p_\perp}^{-3-2n} |{\tilde p_\parallel}|^{-2m} {\rm sign}( \tilde p_\|) + {\tilde q_\perp}^{-3-2n} |{\tilde q_\parallel}|^{-2m} {\rm sign}( \tilde q_\|) \right]
\left[ 1 + {\tilde p_\perp}^{1-n} |{\tilde p_\parallel}|^{1-m} {\rm sign}( \tilde p_\|) + {\tilde q_\perp}^{1-n} |{\tilde q_\parallel}|^{1-m} {\rm sign}( \tilde q_\|) \right]
 \nonumber \\
&&\mbox{} \times 
\delta (s+s_p \tilde p_\perp \tilde p_\parallel + s_q \tilde q_\perp \tilde q_\parallel) 
\delta(1+\tilde p_\parallel+\tilde q_\parallel) d\tilde p_\perp d\tilde q_\perp d\tilde p_\parallel d\tilde q_\parallel \;.  \nonumber
\ea
From this expression, two stationary solutions emerge, namely
\be
n=1 \quad \text{and} \quad m=0 \;,
\ee
and 
\be
n=-3/2 \quad \text{and} \quad m=-1/2 \;.
\ee
The first solution corresponds to the thermodynamic (zero-flux) solution, while the second is the Kolmogorov-Zakharov spectrum, for which the energy flux is finite.

\subsubsection{Cascade direction}
\label{section4}

We now introduce the axisymmetric energy flux,
\be
\partial_t E_k = - \frac{\partial \Pi_\perp (\kpn,k_\parallel)}{\partial \kpn} - \frac{\partial \Pi_\parallel (\kpn, k_\parallel)}{\partial k_\parallel} \;.
\ee
From the kinetic equation (\ref{KEani4}), we write 
\be 
\partial_t E_k = \frac{\epsilon^2 A^2}{384 \nu_\textrm{odd}}  k_\perp^{2n+2} k_\parallel^{2m} I(n,m) \;,
\ee
where $I(n,m)$ is the normalized collisional integral.

Let us now integrate this equation with respect to $k_\perp$ on the one hand, and $k_\|$ on the other hand. Taking the limit $(n,m) \rightarrow \left( -3/2, -1/2 \right)$,  
$\Pi_\perp$ becomes independent of $k_\perp$ and $\Pi_\|$ independent of $k_\|$, such that only one of the two fluxes contributes in each case. Thanks to L'Hospital's rule, we obtain
\ba \label{oddvisc_fluxes}
\Pi_\perp^{KZ} &=& - \frac{\epsilon^2 A^2}{384 \nu_\textrm{odd}}  \frac{1}{2k_\parallel} \frac{\partial I(n,-1/2)}{\partial n}\vert_{n=-3/2} 
\equiv \frac{\epsilon^2 A^2}{384 \nu_\textrm{odd}}  \frac{1}{k_\parallel} I_\perp \;, \\
\Pi_\parallel^{KZ}  &=& - \frac{\epsilon^2 A^2}{384 \nu_\textrm{odd}}  \frac{1}{2\kpn}\frac{\partial I(-3/2,m)}{\partial m}\vert_{m=-1/2} 
\equiv \frac{\epsilon^2 A^2}{384 \nu_\textrm{odd}}  \frac{1}{\kpn} I_\parallel \;,
\ea
where
\ba    
\left( I_\perp \atop I_\| \right) &\equiv& \sum_{s s_p s_q} \int_{\Delta_\perp} \sin \theta_k  
\left( s_q \tilde{q}_\perp - s_p \tilde{p}_\perp \right)^2 \left( s + s_p \tilde{p}_\perp + s_q \tilde{q}_\perp \right)^2 \\
&&\tilde{p}_\perp^{-5/2} \tilde{q}_\perp^{-5/2} |\tilde{p}_\parallel|^{-1/2} |\tilde{q}_\parallel|^{-1/2} 
\left( \tilde{p}_\parallel  \ln \tilde{p}_\perp +  \tilde{q}_\parallel  \ln \tilde{q}_\perp 
\atop \tilde{p}_\parallel \ln \tilde{p}_\parallel + \tilde{q}_\parallel \ln \tilde{q}_\parallel\right) 
 \left(1+ \tilde{p}_\perp^{5/2} |\tilde{p}_\parallel|^{3/2} {\rm sign}(\tilde p_\|) + \tilde{q}_\perp^{5/2} |\tilde{q}_\parallel|^{3/2} {\rm sign}(\tilde q_\|) \right) \nonumber \\
 &&\delta \left( s + s_p \tilde{p}_\perp \tilde{p}_\parallel + s_q \tilde{q}_\perp \tilde{q}_\parallel \right) 
 \delta \left( 1 + \tilde{p}_\| + \tilde{q}_\| \right)  
 d \tilde{p}_\perp d \tilde{q}_\perp d \tilde{p}_\| d \tilde{q}_\| \;. \nonumber
\ea
With this notation, we find the simple relationship for the energy flux ratio
\be
\frac{\Pi_\parallel^{KZ}}{\Pi_\perp^{KZ}} = \frac{k_\parallel}{\kpn} \frac{I_\parallel}{I_\perp} \;.
\ee
Since by assumption we have $\kpn \gg k_\parallel$, this ratio will be small as long as $I_\parallel \sim I_\perp$, meaning that most of the energy transfer in the $k$-space takes place inside the plane orthogonal to ${\bm e}_\|$. A numerical evaluation of the sign of $I_\perp$ shows that the perpendicular energy flux is positive, so the energy cascade is direct, i.e. the energy flows from large scales to small scales.

Finally, let us note that from \eqref{oddvisc_fluxes}, we can also express the constant $A$ as a function of the parameters of the model and the fluxes, which gives the expression (after a rescaling $E\to E/\epsilon^2$ and $\Pi_\perp \to \Pi_\perp/\epsilon^2$)
\be \label{KZscaling1}
E(k_\perp,k_\parallel) = \mathcal{C} \sqrt{\nu_{\rm odd} k_\parallel \Pi_\perp} \, k_\perp^{-3/2} k_\parallel^{-1/2} 
\ee
where $\mathcal{C}=8\sqrt{6/I_\perp}$ is the Kolmogorov constant. Note that in Ref.~\cite{nhwt}, a scaling argument gave the result 
\begin{equation} \label{KZscaling2}
    E(k_\perp,k_\parallel) \sim \sqrt{\varepsilon \nu_\textrm{odd}} \, k_\perp^{-3/2} k_\parallel^{-1/2}
\end{equation}
where $\varepsilon$ is the energy injection rate. This is compatible with our result \eqref{KZscaling1} if we consider that the integral of $\Pi_\perp$ over $k_\|$ (with a cutoff at the dissipation scale) is approximately equal to $\varepsilon$ (since most of the flux is along the perpendicular direction), which gives $\varepsilon\sim k_\| \Pi_\perp$.

\subsubsection{Validity of the weak wave regime and inertial range}

As mentioned at the beginning of this section, the range of validity of weak wave turbulence theory can be estimated by evaluating the ratio of the linear timescale $\tau_{\textrm{lin}}\sim\omega_{\bm{k}}^{-1}$ and the non-linear time scale estimated from the equation of motion \eqref{eq:NS}, $\tau_\textrm{NL} \sim 1/(kv) \sim 1/(k \sqrt{k_\perp k_\parallel E(k_\perp,k_\parallel)})$,
\be
\chi \equiv \frac{\tau_{\textrm{lin}}}{\tau_{\textrm{NL}}} \sim \frac{\sqrt{k_\perp k_\parallel E(k_\perp,k_\parallel)}}{\nu_{\rm odd} k_\|} \sim \varepsilon^{1/4} \nu_\textrm{odd}^{-3/4} k_\perp^{-1/4} k_\parallel^{-3/4} \;,
\ee
where we have used the result \eqref{KZscaling2} for the energy spectrum. This ratio decreases as we go towards the small scales. Thus, weak turbulence remains valid down to the smallest scales, in contrast with rotating (inertial wave) turbulence, where strong turbulence is always recovered below a certain scale \cite{Galtier2003,NazarenkoS2011}.

Let us also comment on the dissipation due to the shear viscosity. Dissipation is negligible as long as the Reynolds number ($R_e$), which in this case is equal to the ratio of the dissipation timescale $\tau_{\rm diss}\sim 1/(\nu k^2)$ to the nonlinear timescale $\tau_\textrm{NL}$, is much larger than $1$. Using again the scaling law \eqref{KZscaling2} to evaluate $\tau_\textrm{NL}$, we find
\be
R_e \sim \frac{\tau_{\rm diss}}{\tau_{\rm NL}} \sim \frac{(\varepsilon \nu_{\rm odd})^{1/4}}{\nu k} \left( \frac{k_\|}{k_\perp} \right)^{1/4} \;.
\ee
Since we have $k_\| \ll k_\perp$, $k$ should be much smaller than $(\varepsilon \nu_{\rm odd})^{1/4}/\nu$ for the dissipation to be negligible. Nevertheless, if the shear viscosity $\nu$ is small enough, it is in principle possible for the dissipation to only affect the small scales, allowing for the emergence of an inertial range.

\subsection{Comparison with the numerical results}

To complement the analytical study presented here, we have also performed direct numerical simulations of the model Eqs.~\eqref{eq:NS}. A precise discussion of the method and of the obtained results are provided in Ref.~\cite{nhwt}. To maximize the size of the inertial range, we force the system by applying Gaussian noise in a narrow band of wavenumbers at large scales, and we replace the viscosity term by a hyperviscosity term of the form $\nu_\alpha \Delta^{\alpha} \bm{v}$. By increasing the odd viscosity parameter $\nu_{\rm odd}$, we are able to go from the critical balance regime $\chi\sim \mathcal{O}(1)$, to the weak wave turbulence regime $\chi\ll 1$. We check that we are in the correct regime by plotting the ratio $\chi$, as well as the spatio-temporal energy spectrum (which is concentrated around the dispersion relation in the weak wave turbulence regime, while in the critical balance regime a wider range of modes are excited).

In both regimes, we have plotted the kinetic energy spectrum as a function of $k_\perp$ (our simulations did not allow us to test the scaling prediction for the parallel direction). In the critical balance regime, the result is compatible with the expected scaling \mbox{$E(k_\perp, k_\parallel)\sim k_\perp^{-5/3} k_\parallel^{-1}$} \cite{Higdon1984,GS95,Nazarenko2011,Oughton2020,Zhou2021,Alexakis2018}. In the weak wave turbulence regime, we find a good agreement with the KZ solution derived above, \mbox{$E(k_\perp, k_\parallel)\propto  k_\perp^{-3/2}$}.

\section{A toy model for odd wave turbulence in odd elastic materials} \label{sec:odd_solid}

\subsection{Incompressible odd elastic solid in 1D} \label{sec:odd_solid_construct}

In this section, we introduce a model of incompressible (shear-wave) odd elastic solid in 1D, which is summarized at the beginning of Sec.~\ref{odd_elastic_model}.

\subsubsection{Introduction of the model}
\label{intro_model}

In this second part of the paper, we study the propagation of one-dimensional nonlinear shear waves in incompressible chiral active solids. 
The evolution of the displacement field $\bm{u}(t,z)$ (where $z$ is the direction of propagation) is given by
\begin{equation}
    \rho_0 \partial_t^2 \bm{u} + \Gamma \partial_t \bm{u} = \bm{\nabla}\cdot\bm{\sigma}
\end{equation}
in which $\bm{\sigma}$ is the stress tensor and $\rho_0$ is the density of the unperturbed medium.
This simplifies into $[\bm{\nabla}\cdot\bm{\sigma}]_i = \partial_z \sigma_{i z}$ as fields only depend on $z$.
As incompressibility imposes $\partial_z u_z = 0$, we now consider only transverse displacements $\bm{u}(t,z) \simeq [u_x(t,z), u_y(t,z)]^T$.
The relation between stress and deformation gradients takes the general form
\begin{equation}
    \sigma_{i z} = 
    A^{(1)}_{i j} \partial_z u_j 
    + 
    A^{(2)}_{i j k} (\partial_z u_j) (\partial_z u_k)
    +
    A^{(3)}_{i j k \ell} (\partial_z u_j) (\partial_z u_k) (\partial_z u_\ell)
    + \mathcal{O}(u^4) \;.
\end{equation}
Imposing rotation invariance about the $z$ axis leads to constraints on the $A^{(n)}$ tensors: for instance, the most general rank two tensor compatible with this symmetry in 2D is a linear combination of $\delta_{ij}$ (the identity tensor) and $\epsilon_{ij}$ (the fully antisymmetric tensor).
Similarly, $A^{(2)} = 0$ (note that this removes second-order nonlinearities).
Taking into account the permutation invariance with respect to the indices $j$, $k$, $\ell$, we can also find the most general $A^{(3)}$ is a linear combination of 
$\frac{1}{3} (\delta_{i  \ell} \delta_{j  k}+\delta_{i  k} \delta_{j  \ell}+\delta_{i  j} \delta_{k  \ell})$ and 
$\frac{1}{3} (2 \epsilon_{i  j} \delta_{k  \ell}+\epsilon_{i  k} \delta_{j  \ell}+\delta_{i  k} \epsilon_{j  \ell})$. 
Putting it all together yields
\begin{equation}
    \label{eom_incomp_soft_solid}
    \rho_0 \partial_t^2 \bm{u} + \Gamma \partial_t \bm{u} = \partial_z \big[(\mu + \mu_{\text{odd}} \bm{\epsilon}) \partial_z \bm{u} + (\mu^{\text{NL}} + \mu^{\text{NL}}_{\text{odd}} 
    \bm{\epsilon}) \lVert \partial_z \bm{u} \rVert^2 \partial_z \bm{u} \big] + \bm{f} \quad , \quad \boldsymbol{\epsilon} = \begin{pmatrix} 0 & 1 \\ -1 & 0 \end{pmatrix}
\end{equation}
in which $\rho_0$ is the density of the unperturbed medium, $\Gamma$ is a linear damping coefficient, $\mu$ is the shear modulus, and $\mu^{\text{NL}}$ is a nonlinear shear modulus, all of which arise in standard solids.
In addition, the linear and nonlinear odd elastic moduli $\mu_{\text{odd}}$ and $\mu^{\text{NL}}_{\text{odd}}$ induce a chiral coupling between the two polarizations $u_x$ and $u_y$ \cite{Scheibner2020b,oddreview}. As in the case of the odd fluid, we also add a stochastic forcing $\bm{f}$.

\subsubsection{Overdamped limit and wave propagation} 

In order to allow for the propagation of waves, we consider the overdamped limit,
\begin{equation}
    \label{eom_incomp_soft_solid_overdamped}
    \partial_t \bm{u} = (G + G_{\text{odd}} \bm{\epsilon}) \partial_z^2 \bm{u} + (G^{\text{NL}} + G^{\text{NL}}_{\text{odd}} 
    \bm{\epsilon}) \partial_z\big[\lVert \partial_z \bm{u} \rVert^2 \partial_z \bm{u} \big] + \bm{g}
\end{equation}
in which we introduce $G_i = \mu_i/\Gamma$ for each of the four elastic moduli, and $\bm{g}=\bm{f}/\Gamma$. 
At the linear level, Eq.~\eqref{eom_incomp_soft_solid_overdamped} admits wave-like solutions. Substitution of a solution of the form $\exp(i(\omega t - k z))$ yields
\begin{equation}
    \omega = i k^2G \pm k^2 G_\textrm{odd} \;,
\end{equation}
which indicates harmonic solutions that are damped at a rate $k^2G$. For strong odd elasticity $G_\textrm{odd} \gg G$, this yields the dispersion relation
\begin{equation} \label{disp_rel}
    \omega = \pm k^2 G_\textrm{odd} \;.
\end{equation}
The derivation of \eqref{eom_incomp_soft_solid_overdamped} from \eqref{eom_incomp_soft_solid} is detailed in the Appendix~\ref{app:overdamped}. This approximation is valid as long as the timescale $\rho_0/\Gamma$ is much smaller than the smallest relevant timescale for the system, namely $\omega^{-1}$, which gives the condition 
\be \label{k_inertia}
k \ll k_{\rm inertia} = \sqrt{\frac{\Gamma}{\rho_0 G_{\rm odd}}} = \frac{\Gamma}{\sqrt{\rho_0 \mu_{\rm odd}}} \;.
\ee
Thus, the role of inertia can be neglected on scales larger than $k_{\rm inertia}^{-1}$, which is the regime we will focus on from now on. Note that on scales where it becomes relevant, the inertia modifies the dispersion relation, adding an imaginary contribution. It may thus play a dissipative role in terms of wave propagation.

Since the system admits wave-like solutions that are subject to a weak non-linear interaction, we may anticipate wave turbulence, similar to the odd viscous wave turbulence discussed in Section \ref{sec:odd_fluid}. In this analogy, the regular elasticity $G$ plays the role of the regular viscosity $\nu$, being a dissipative force, while the odd elasticity $G_\textrm{odd}$ gives rise to waves with a similar dispersion relation to what is found for odd viscosity $\nu_\textrm{odd}$. Note that anisotropy is absent here because the model has only a single spatial direction. Another notable difference is that the non-linearity is cubic, which corresponds to a four-wave interaction. This is important since, as we will see, even-order interactions may lead to dual cascades, where one conserved quantity flows towards the small scales while another one flows towards the large scales. Finally, and perhaps most importantly, contrary to odd viscosity, odd elasticity does not conserve energy. There are however other conserved quantities which emerge in this system, as we will discuss below. These are the quantities that will be relevant for wave turbulence, playing the same role as the energy in the odd fluid.

\subsubsection{Shear elasticity and dissipation} 

An important requirement for the application of weak wave turbulence theory is the existence of an inertial range where forcing and dissipation are negligible. However, in the present case, the ratio of the dissipation timescale $\tau_{\rm diss}\sim 1/(G k^2)$ due to shear elasticity with the linear timescale $\tau_\textrm{lin} \sim \omega^{-1} \sim 1/(G_{\rm odd}k^2)$ is independent of $k$. This suggests that the role of the shear elasticity is the same at all scales. We will confirm this at the end of this section by also comparing the dissipation timescale with the non-linear timescale. This implies that, if we want the dissipation due to the shear elasticity to be negligible, which can be achieved by setting $G\ll G_{\rm odd}$, then it will be negligible at all scales. Therefore, it cannot play the role of a small scale (or large scale) dissipation, as was the case for the shear viscosity. Instead, the inertial range will be determined by other sources of dissipation not included in \eqref{eom_incomp_soft_solid}\footnote{We note that, if no other source of dissipation is present in the system apart from a small shear elasticity, then the wave action will be dissipated over a very large window of wavenumbers, and weak wave turbulence will still be approximately valid when considering a small subpart of this window close to the injection scale, but this is generally not what will happen in practice in a realistic system.}. For instance, inertia could dissipate wave action at small scales, while the large scale dissipation could be due to boundary effects. A similar situation occurs in Hall magnetohydrodynamics turbulence, where the Hall term also involves second order spatial derivatives, and thus has the same scale dependence as the viscous term \cite{HallMHD}.

In the rest of the paper, we will assume that both the linear and non-linear shear elasticity can be neglected, and we will instead consider an equation of motion of the form 
\begin{equation}
    \label{eom_odd_solid_true}
    \partial_t \bm{u} = G_{\text{odd}} \bm{\epsilon} \partial_z^2 \bm{u} + G^{\text{NL}}_{\text{odd}} 
    \bm{\epsilon} \partial_z\big[\lVert \partial_z \bm{u} \rVert^2 \partial_z \bm{u} \big] + \bm{\mathcal{D}} + \bm{g}
\end{equation}
where $\bm{\mathcal{D}}$ is a dissipation process, which is assumed to be non-zero only at large scales $k<k_-$ and at small scales $k>k_+$, and $\bm{g}$ is a stochastic driving force, which takes non-zero values only for a small range of wavenumbers around some intermediate forcing scale $k_f$, such that $k_-\ll k_f \ll k_+$. 
Below we will study the behavior of the system inside the two inertial ranges $k_- < k <k_f$ and $k_f<k<k_+$, where both forcing and dissipation are negligible. We thus simply assume that the two inertial ranges are large enough so that weak wave turbulence theory may apply, and we will not include these terms explicitly in our equations. 

As a final comment, let us stress that the function $\bm{\mathcal{D}}$ does not describe the dissipation of energy, but rather of the two relevant conserved quantities that will be introduced below, which we will call the wave action and the odd energy. It may therefore involve different processes than the ones that we usually have in mind when we think about dissipation.

\subsubsection{Possible experimental realizations}

Equation \eqref{eom_incomp_soft_solid} has been obtained from symmetry. Below we give an example of a microscopic model which reduces to Eq.~\eqref{eom_incomp_soft_solid} upon coarse-graining, namely a one-dimensional chain where particles confined in fixed $z$ planes and free to move within the plane are connected to their neighbors by both normal and odd springs, with a rest length inferior to the distance between the planes. 
This could be implemented using robotic feedback loops, which have already been used to realize materials with odd elastic responses~\cite{Chen2021,Veenstra2025}. 
Passive incompressible soft solids also include artificial gels and certain biological tissues~\cite{Zabolotskaya2004,Catheline2003a,Catheline2003b,Jacob2007,Destrade2019}. 
These could be combined with mechanochemical feedbacks or electroactive elastomers \cite{Apsite2021,Mirfakhrai2007,Li2021,Liu2020,Chung2025} or with a miniaturized version of robotic feedback loops, implemented for instance using the methodology described in Refs.~\cite{Miskin2020,Reynolds2022,Liu2024} where micron-scale robotic metasheets 
are experimentally demonstrated, that could be embedded in a gel realizing the passive soft solid.

Note that in Eq.~\eqref{eom_incomp_soft_solid}, dissipation is modeled by a linear damping $\Gamma$, like in the case of vibrating plates \cite{Dyachenko2004,During2006} where this captures the basic experimental phenomenology, although in general a scale-dependent damping $\Gamma(k)$ may be present \cite{Miquel2011,Miquel2014,Humbert2013,Hassaini2019}.
In addition, a viscous dissipation term $\eta \partial_t \partial_z^2 \bm{u}$ can be added to Eq.~\eqref{eom_incomp_soft_solid}, where $\eta$ is the shear viscosity of the gel; this has been used to model nonlinear shear waves in experiments on soft gels \cite{Zabolotskaya2004,Catheline2003a,Catheline2003b,Jacob2007,Destrade2019}.
In both cases, we expect that this will induce an additional regime, the properties of which can be determined in the same way as done below for odd elastic waves by using the full dispersion relation.
Here, we focus on the regime where these effects can be neglected; for instance, the viscous dissipation can be neglected when $\eta \omega \ll \mu$.

\subsubsection{Derivation from a microscopic model}

\begin{figure}
    \centering
    \includegraphics[width=0.6\linewidth,trim={2.5cm 6.2cm 2.4cm 6cm},clip]{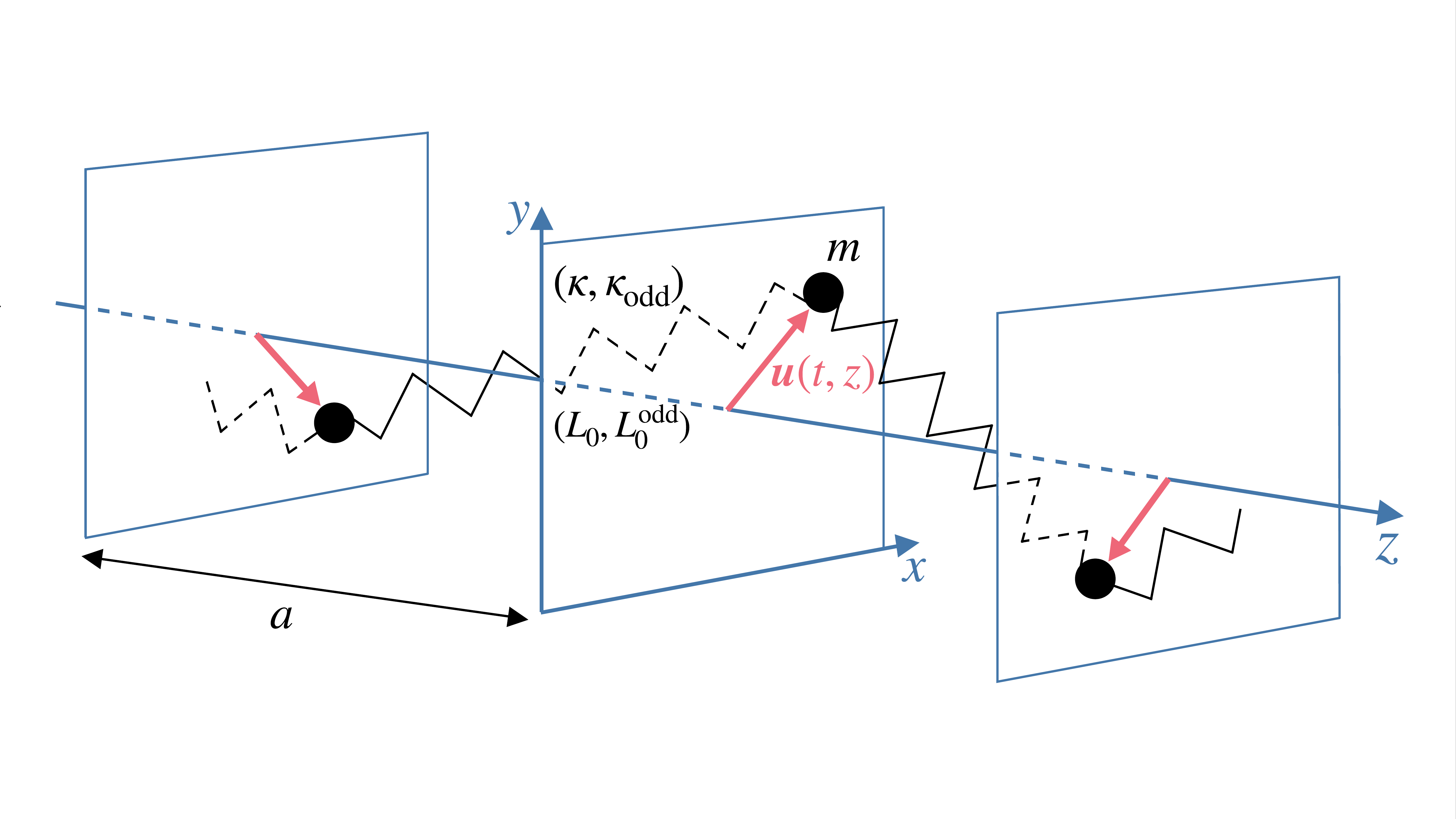}
    \caption{
    \textbf{Microscopic model of a quasi-one-dimensional shear-only odd solid.}
    Sketch of a microscopic model which can be described by the equation \eqref{elastic_model} in the continuous limit. It consists in a chain of particles connected by both normal and odd springs and constrained to move in a plane orthogonal to the direction of the chain.}
    \label{fig:odd_elastic_sketch}
\end{figure}

The equation \eqref{eom_incomp_soft_solid} can be obtained by taking the continuous limit of a microscopic model, such as the one depicted in Fig.~\ref{fig:odd_elastic_sketch}. Consider a chain of particles of mass $m$. Each particle is constrained to move inside a plane $(x,y)$ orthogonal to the direction $z$ of the chain. Two neighboring planes are separated by a distance $a$. Neighboring particles are connected both by a normal spring, with spring constant $\kappa$ and rest length $L_0$, and by an odd spring, with spring constant $\kappa_{\rm odd}$ and rest length $L_0^{\rm odd}$. We choose $L_0<a$ and $L_0^{\rm odd}\neq a$, such that the system is pre-stressed in order to have linear shear elasticity (and such that the normal springs tend to align the particles). Denoting $\bm{\tilde u}^j(t) \simeq [\tilde u_x^j(t), \tilde u_y^j(t)]^T$ the displacement of the particle with label $j$ inside the $(x,y)$ plane, the force exerted on the particle $j$ by the particle $j-1$ reads
\be
\bm{F}_{j-1\to j} = -\kappa(L_{j-1,j} - L_0) \, \bm{e_{j-1,j}}^\perp - \kappa_{\rm odd} (L_{j-1,j} - L_0^{\rm odd}) \, \bm{\epsilon} \, \bm{e_{j-1,j}}^\perp
\ee
where $\bm{\epsilon}$ again denotes the fully antisymmetric tensor defined in \eqref{eom_incomp_soft_solid}, and
\bea
&&L_{j-1,j} = \sqrt{a^2 + \|\bm{\tilde u}^j - \bm{\tilde u}^{j-1} \|^2} \;, \\
&&\bm{e_{j-1,j}}^\perp = \frac{\bm{\tilde u}^j - \bm{\tilde u}^{j-1}}{L_{j-1,j}} \;.
\eea
Assuming in addition that each particle is subject to a friction coefficient $\gamma$, we have the equation of motion
\be \label{discrete_eom}
m \partial_t^2 \bm{\tilde u}^j + \gamma \partial_t \bm{\tilde u}^j = \bm{F}_{j-1\to j} + \bm{F}_{j+1\to j} \;.
\ee
We now take the continuous limit by introducing a continuous function $\bm{\tilde u}(t,z)$ such that $\bm{\tilde u}(t,\tilde z=ja)=\bm{\tilde u}^j(t)$. We assume that this function varies on a scale $\ell$ which is very large compared to both the lattice spacing $a$ and the displacements $\|\bm{\tilde u}\|$, which allows us to perform a Taylor expansion. Before that, let us rescale the function $\bm{\tilde u}$ by defining $\bm{\tilde u}(t,\tilde z)=a  \bm{u}(t,z=\tilde z/\ell)$, such that the function $\bm{u}(t,z)$ is now of order 1 and varies on a scale of order 1. This leads to
\be
\bm{u}^j - \bm{u}^{j\pm 1} \simeq -a \left(\pm \frac{a}{\ell} \partial_z \bm{u} + \frac{1}{2} \frac{a^2}{\ell^2} \partial_z^2 \bm{u} \pm \frac{1}{6} \frac{a^3}{\ell^3} \partial_z^3 \bm{u} + \frac{1}{24} \frac{a^4}{\ell^4} \partial_z^4 \bm{u} + \mathcal{O}(a^5/\ell^5) \right) \;.
\ee
The spring length then reads
\be
L_{j\pm 1,j} \simeq a \left( 1 + \frac{1}{2} \frac{a^2}{\ell^2} \|\partial_z \bm{u}\|^2 \pm \frac{1}{4} \frac{a^3}{\ell^3} \partial_z \|\partial_z \bm{u}\|^2 + \mathcal{O}(a^4/\ell^4) \right) \;.
\ee
Inserting into \eqref{discrete_eom}, we obtain
\bea \label{continuous_eom}
m \partial_t^2 \bm{u} + \gamma \partial_t \bm{u} &\simeq& \kappa \left[ \frac{a^2}{\ell^2} \big(1-\frac{L_0}{a}\big) \partial_z^2 \bm{u} + \frac{1}{12} \frac{a^4}{\ell^4} \big(1-\frac{L_0}{a}\big) \partial_z^4 \bm{u} +\frac{a^4}{\ell^4} \frac{L_0}{2a} \partial_z [ \|\partial_z \bm{u}\|^2 \partial_z \bm{u} ] \right] \\
&& + \kappa_{\rm odd} \left[ \frac{a^2}{\ell^2} \big(1-\frac{L_0^{\rm odd}}{a}\big) \bm{\epsilon} \partial_z^2 \bm{u} + \frac{1}{12} \frac{a^4}{\ell^4} \big(1-\frac{L_0^{\rm odd}}{a}\big) \bm{\epsilon} \partial_z^4 \bm{u} + \frac{a^4}{\ell^4} \frac{L_0^{\rm odd}}{2a} \bm{\epsilon} \partial_z [ \|\partial_z \bm{u}\|^2 \partial_z \bm{u} ] \right] + \mathcal{O}(a^5/\ell^5) \nn \;.
\eea
In order to recover \eqref{eom_incomp_soft_solid} we only need to make one final assumption, namely that $|a-L_0|\ll a$ and $|a-L_0^{\rm odd}|\ll a$. With this assumption, the term containing $\partial_z^4 \bm{u}$ is negligible. In addition, if we assume the scaling
\bea
&&1-\frac{L_0}{a} = \frac{a^2}{\ell^2} \Delta \;, \\
&&1-\frac{L_0^{\rm odd}}{a} = \frac{a^2}{\ell^2} \Delta_{\rm odd} \nn
\eea
where $\Delta$ and $\Delta_{\rm odd}$ are both of order 1, then the linear and non-linear term are of the same order. This finally leads to (the factors $L_0/a$ and $L_0^{\rm odd}/a$ can both be approximated to 1 in the non-linear term)
\be \label{continuous_eom_final}
m \partial_t^2 \bm{u} + \gamma \partial_t \bm{u} \simeq \frac{a^4}{\ell^4} \left[ (\kappa \Delta + \kappa_{\rm odd} \Delta_{\rm odd} \bm{\epsilon} ) \partial_z^2 \bm{u} + ( \frac{\kappa}{2} + \frac{\kappa_{\rm odd}}{2} \bm{\epsilon}) \partial_z [ \|\partial_z \bm{u}\|^2 \partial_z \bm{u} ] \right] + \mathcal{O}(a^5/\ell^5) \;.
\ee
Defining
\be
\rho_0=m/a \;, \quad \Gamma=\gamma/a \;, \quad \mu=\kappa\Delta a^3/\ell^4 \;, \quad \mu_{\rm odd} =\kappa_{\rm odd}\Delta_{\rm odd} a^3/\ell^4 \;, \quad \mu^{\rm NL}=\kappa a^3/(2\ell^4) \;, \quad \mu_{\rm odd}^{\rm NL} =\kappa_{\rm odd} a^3/(2\ell^4) \;,
\ee
we finally recover \eqref{eom_incomp_soft_solid}. Note that we have assumed that the smallest variations of $\bm{u}$ are on a scale $\sim \ell\gg a$, thus this description is only valid for wavenumbers $k \ll a^{-1}$.

\subsection{Odd elastic solid model: canonical form and conserved quantities}
\label{odd_elastic_model}

\subsubsection{A simple model of odd elastic solid}

In the remainder of the paper, we focus on the model of a one-dimensional shear-wave odd elastic solid constructed in Sec.~\ref{sec:odd_solid_construct} where the deformation field $\bm{u}(t, \bm{r})$ follows
\begin{equation}
    \label{eom_odd_solid_true}
    \partial_t \bm{u} = G_{\text{odd}} \bm{\epsilon} \partial_z^2 \bm{u} + G^{\text{NL}}_{\text{odd}} 
    \bm{\epsilon} \partial_z\big[\lVert \partial_z \bm{u} \rVert^2 \partial_z \bm{u} \big] + \bm{\mathcal{D}} + \bm{g}
\end{equation}
where $\bm{\mathcal{D}}$ describes the dissipation of wave action and odd energy, while $\bm{g}$ is a stochastic driving force. We refer to Sec.~\ref{sec:odd_solid_construct} for details and Sec.~\ref{numerical_simulations_odd_solid} for the particular choices of $\bm{\mathcal{D}}$ and $\bm{g}$ used in numerical simulations.

\subsubsection{Equation in canonical form and conserved quantities}

The existence of waves which interact via a weak non-linearity is not the only ingredient required for the development of wave turbulence: it also requires the existence of one or several conserved quantities which can cascade through scales. This is where the main difference between odd viscous fluids and odd elastic solids arises. Indeed, contrary to odd viscosity, odd elasticity does not conserve energy. For our model, the total energy reads
\be \label{elastic_energy}
\mathcal{E} = \int dz \left[ \frac{1}{2} \rho_0 \|\partial_t \bm{u} \|^2 + \frac{1}{2} \mu \|\partial_z \bm{u} \|^2 + \frac{1}{4} \mu^{\rm NL} \|\partial_z \bm{u} \|^4 \right] \;.
\ee
Since the odd elastic force cannot be written in a conservative form, it does not appear in the energy. The time evolution of the energy thus reads (omitting boundary terms)
\bea
\frac{d\mathcal{E}}{dt} &=& \int dz \; \partial_t \bm{u} \cdot \left[ \rho_0 \partial_t^2 \bm{u} - \mu \partial_z^2 \bm{u} - \mu^{\rm NL} \partial_z [ \|\partial_z \bm{u} \|^2 \partial_z \bm{u} ] \right] \\
&=& \int dz \; \partial_t \bm{u} \cdot \left[ \mu_{\rm odd} \bm{\epsilon} \partial_z^2 \bm{u} + \mu^{\rm NL}_{\rm odd} \bm{\epsilon} \partial_z [ \|\partial_z \bm{u} \|^2 \partial_z \bm{u}] - \Gamma \partial_t u \right]
\nn
\eea
where in the second line we have used the equation of motion \eqref{eom_incomp_soft_solid}. We clearly see that when the odd elastic coefficients are non-zero, the energy is not conserved\footnote{More generally, another way to see this is to show that one can find a cycle of deformations for which the total work is non-zero \cite{oddreview}.} (of course this is all the more true in the presence of damping). Obviously, a turbulent cascade of energy as the one we have evidenced in the odd fluid case is therefore not possible. However, we will now show that if we consider the overdamped regime \eqref{eom_incomp_soft_solid_overdamped}, and if we assume in addition that the odd elastic effects dominate over the even effects, i.e. $G_\textrm{odd} \gg G$ and $G_\textrm{odd}^\textrm{NL} \gg G^\textrm{NL}$, different conserved quantities emerge. 

Before that, let us note that the evolution equation for \eqref{eom_incomp_soft_solid_overdamped} is not written in a canonical form. If we want to apply weak wave turbulence theory to this system, it will be much more convenient to rewrite this equation in a canonical form. For this, let us first multiply this equation by $i$ to rewrite it as (from now on we set the coefficients $G$ and $G^{\rm NL}$ to zero)
\be \label{elastic_model}
i\partial_t \mathbf{u} = -G_{\rm odd} \boldsymbol{\tilde \epsilon} \partial_z^2 \mathbf{u} - G^{\text{NL}}_{\text{odd}} \boldsymbol{\tilde \epsilon} \partial_z [\|\partial_z \mathbf{u}\|^2 \partial_z \mathbf{u}] \quad , \quad \boldsymbol{\tilde \epsilon} = -i \boldsymbol{\epsilon} = \begin{pmatrix} 0 & -i \\ i & 0 \end{pmatrix} \;,
\ee
where the matrix $\boldsymbol{\tilde \epsilon}$ is now Hermitian. Thus, it can be diagonalized as
\be
\boldsymbol{\tilde \epsilon} = \mathbf{P} \begin{pmatrix} 1 & 0 \\ 0 & -1 \end{pmatrix} \mathbf{P}^\dagger \quad , \quad \mathbf{P} = \frac{1}{\sqrt{2}} \begin{pmatrix} 1 & 1 \\ i & -i \end{pmatrix} \;,
\ee
where $\mathbf{P}$ is a unitary matrix. This suggests a change of variables
\be
\Psi(t,z) = \sqrt{2} \bm{P}^\dagger \bm{u}(t,z) = \begin{pmatrix} u_x(t,z) - iu_y(t,z) \\ u_x(t,z) + iu_y(t,z) \end{pmatrix} \equiv \begin{pmatrix} \psi(t,z) \\ \psi^*(t,z) \end{pmatrix} \;.
\ee
The equation of motion can thus be written in terms of a single complex scalar variable $\psi(t,z)$,
\be \label{elastic_model_psi}
i\partial_t \psi = -G_{\rm odd} \partial_z^2 \psi - G^{\text{NL}}_{\text{odd}} \partial_z [|\partial_z \psi|^2 \partial_z \psi]
\ee
where we have used that $|\partial_z \psi|^2 = \|\partial_z \mathbf{u}\|^2$. From this equation, one can now easily define a Hamiltonian as
\be \label{Hamiltonian2}
\mathcal{H} = \int dz \, [G_{\rm odd} |\partial_z \psi|^2 + \frac{1}{2} G_{\rm odd}^{NL} |\partial_z \psi|^4]
= \int dz \, [G_{\rm odd} \|\partial_z \mathbf{u}\|^2 + \frac{1}{2} G_{\rm odd}^{\rm NL} \|\partial_z \mathbf{u}\|^4]
\ee
such that
\be \label{elastic_model_psi_canonical}
i\partial_t \psi = \frac{\delta \mathcal{H}}{\delta \psi^*} \;.
\ee
The expression \eqref{Hamiltonian2} has a similar form to the usual elastic energy in \eqref{elastic_energy}, but with the odd elastic coefficients $G_{\rm odd}$ and $G_{\rm odd}^{\rm NL}$ playing the role of the shear elastic moduli $\mu$ and $\mu_{\rm NL}$. Note that writing the equation \eqref{elastic_model_psi} in terms of $\mathbf{u}$ gives
\be
\partial_t \mathbf{u} = -\boldsymbol{\epsilon} \frac{\delta \mathcal{H}}{\delta \mathbf{u}}
\ee
which, as mentioned above, is not in canonical form.

To derive the wave kinetic equation for this model, we will need to work in Fourier space (assuming as in the first part an infinite size system). Introducing $A_k(t) = \int \frac{dz}{2\pi} \, e^{-ikz} \psi(t,z)$, the equation of motion reads
\be \label{Hamilton_eq_a}
i\partial_t A_k = \frac{\delta \mathcal{H}}{\delta A_k^*}  \;,
\ee
with the same Hamiltonian $\mathcal{H}$ as before, which, introducing $\omega_k=G_{\rm odd}k^2$ and $T_{12}^{34}=2\,G^{\text{NL}}_{\text{odd}} \, k_1k_2k_3k_4$, reads in terms of the variable $A_k$
\be \label{Hamiltonian1}
\mathcal{H} = \int dk \, \omega_k \, |A_k|^2 + \frac{1}{4} \int dk_1dk_2dk_3dk_4 \, T_{12}^{34} \, A_1^* A_2^* A_3 A_4 \delta(k_1+k_2-k_3-k_4) \;.
\ee
\\

\noindent\textbf{Conserved quantities.}
From the equation written in canonical form \eqref{elastic_model_psi_canonical}, it is now easy to show that the following quantities are conserved:
\\

\noindent (i) the wave action 
\be \label{defWaveAction2}
\mathcal{N}=\int dz \|u\|^2=\int dz |\psi|^2=\int dk\, |A_k|^2 \;,
\ee

\noindent (ii) the Hamiltonian (or { odd energy}) $\mathcal{H}$, defined in \eqref{Hamiltonian2} (or equivalently \eqref{Hamiltonian1}),
\\

\noindent (iii) the { odd momentum} 
\be\label{defPmomentum}
\mathcal{P} = \frac{1}{2} \int dz \, [u_y \partial_z u_x - u_x \partial_z u_y] = \frac{i}{2} \int dz \, [\psi \partial_z \psi^* - \psi^* \partial_z \psi] = \int dk \, k |A(k)|^2\;.
\ee
This last quantity is zero if the action spectrum is even in $k$, and thus it will not play a role here. 
\\

\noindent\textit{Proof:} For the wave action (i), we have
\be
\partial_t \mathcal{N} = \int dz \left( -i \psi^* \frac{\delta \mathcal{H}}{\delta \psi^*} + i \psi \frac{\delta \mathcal{H}}{\delta \psi} \right) = 2 \, {\rm Im} \int dz \, \psi^* \frac{\delta \mathcal{H}}{\delta \psi^*} = 0 \;. \nn
\ee
For the Hamiltonian (ii), we write
\be
\partial_t \mathcal{H} = \frac{\delta \mathcal{H}}{\delta \psi} \partial_t \psi + \frac{\delta \mathcal{H}}{\delta \psi^*} \partial_t \psi^* = -i \frac{\delta \mathcal{H}}{\delta \psi} \frac{\delta \mathcal{H}}{\delta \psi^*} +i \frac{\delta \mathcal{H}}{\delta \psi^*} \frac{\delta \mathcal{H}}{\delta \psi} = 0 \;.
\ee
Finally, for the momentum (iii) the proof is similar to (i).
\\

We thus have two conserved quantities (the wave action and the odd energy) which emerge as a consequence of the odd elasticity, and which may potentially play the same role as the energy in the odd viscous fluid, allowing for a turbulent cascade to develop.

\subsection{Weak wave turbulence theory} 
\label{sec:WWTelastic}

\subsubsection{Wave turbulence phenomenology and overview}

We have now shown that our model possesses all the ingredients for wave turbulence: (i) the existence of wave-like solutions that are subject to a weak non-linear interaction, and (ii) several conserved quantities, which in this case are emergent quantities specific to this system. Focusing on the limit of weak non-linearity, i.e. $G_{\rm odd}^{\rm NL} \ll G_{\rm odd}$, we can therefore apply the framework of weak wave turbulence theory to our system. The main quantity of interest in this case will be the wave action spectrum, defined as (assuming statistically homogeneous turbulence)
\be \label{defn}
n(k) \delta(k-k') = \langle A_k A_{k'}^{*} \rangle \;.
\ee
Denoting $\mathbf{\hat u}_k(t)$ the Fourier transform of $\bm{u}(t,z)$, we have the relation $\|\mathbf{\hat u}_k\|^2=\frac{1}{2}(|A_k|^2+|A_{-k}|^2)$. We thus also introduce
\be \label{defN}
N(k) = \frac{1}{2}\left(n(k)+n(-k)\right) \,, \quad \text{such that} \quad N(k) \delta(k-k') = \langle \mathbf{\hat u}_k^\dagger \mathbf{\hat u}_{k'} \rangle \,.
\ee
Starting from the equation of motion in canonical form \eqref{Hamilton_eq_a}, our first goal will be to derive the wave kinetic equation satisfied by $n(k)$ in the weak regime, before looking for stationary solutions of this equation. For this part of the paper, we will follow the { random phase approximation} method, as presented in \cite{Zakharov1992}, for the derivation of the WKE. This method is slightly different from the multiple time scale method used above for the odd fluid. In particular, we do not explicitly introduce a small parameter $\epsilon$, but the coefficient in front of the nonlinearity is assumed to be small, i.e. $G_{\rm NL}^{\rm odd}/G_{\rm NL}=\mathcal{O}(\epsilon)$ with $\epsilon\ll 1$.

\begin{figure}
    \centering
    \includegraphics[width=0.45\linewidth]{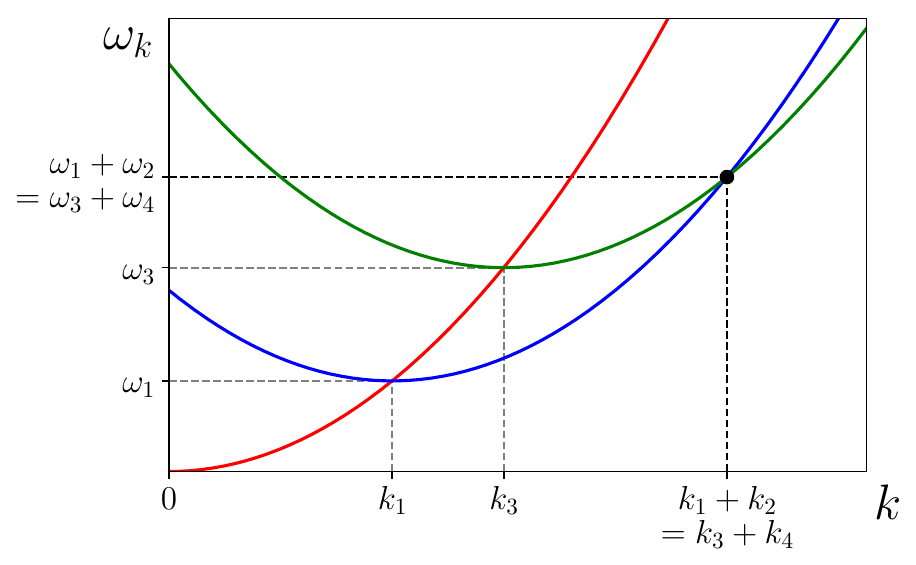}
    \caption{Graphical representation of the solutions of the set of equations \eqref{4wave_manifold} defining the resonant manifold for a four-wave interaction. For a given choice of $k_1$ and $k_3$, the solutions are given by the intersections between the blue and green curves. For a convex dispersion relation, we see that there is a unique solution (black dot), which corresponds to the trivial case $k_2=k_3$ and $k_4=k_1$.}
    \label{fig:dispersion_relation}
\end{figure}

There is however a crucial point that one should note in order to derive the correct kinetic equation for this model. Given that the interaction term is cubic, one would expect weak turbulence to be described by a four-wave interaction. However, this interaction is actually non-resonant since the system of equations
\bea \label{4wave_manifold}
&&k_1+k_2 = k_3+k_4 \;, \\
&&k_1^2+k_2^2 = k_3^2+k_4^2 \;, \nn
\eea
only admits trivial solutions. More generally, there are no four-wave interactions in 1d for systems with a convex dispersion relation, $\omega\propto k^\alpha$ with $\alpha>1$. This can be understood through a simple graphical representation, as shown in Figure~\ref{fig:dispersion_relation}. On this plot, the red curve represents both $\omega_1$ as a function of $k_1$ and $\omega_3$ as a function of $k_3$. For a given choice of $k_1$ and $k_3$, we then plot $\omega_2$ as a function of $k_2$ (in blue) and $\omega_4$ as a function of $k_4$ (in green), with the origin placed at $(k_1,\omega_1)$ and $(k_3,\omega_3)$ respectively. The points where these two curves intersect correspond to the solutions of the set of equations \eqref{4wave_manifold}. For a convex dispersion relation (such as $\omega_k\propto k^2$), we see that there is a unique solution (marked by a black dot), which corresponds to the trivial case $k_2=k_3$ and $k_4=k_1$. Thus, there does not exist any solution which could generate a transfer of wave action between scales.

This means that we first need to perform a canonical transformation to eliminate the non-resonant four-wave interaction, following the method presented in \cite{Zakharov1992,Krasitskii_1994}. This method has previously been applied to eliminate non-resonant four-wave interactions in the context of non-linear optics \cite{Laurie2012} and Kelvin waves in superfluid turbulence \cite{Boffetta2009,Laurie2010}. We expect this process to generate a six-wave interaction (the conservation of the wave action prevents the existence of odd-number interactions). The resonant manifold will thus be instead described by the equations
\bea \label{6wave_manifold}
&&k_1+k_2+k_3 = k_4+k_5+k_6 \;, \\
&&k_1^2+k_2^2+k_3^2 = k_4^2+k_5^2+k_6^2 \;, \nn
\eea
which now have non-trivial solutions. We note that there would be other ways to arrive at the same result. In particular, using the multiple time scale technique would in theory allow us to bypass the need for a canonical transformation, while still leading to the same six-wave interactions \citep{Newell1968}.
\\

In the rest of this section, we present the full weak-wave turbulence computation for our odd elastic solid. We first perform the canonical change of variable required to eliminate the non-resonant four-wave interaction. From the obtained equation of motion, we then derive the WKE satisfied by $n(k)$. Performing a KZ transformation, we then obtain two Kolmogorov-Zakharov solutions for the WKE, corresponding respectively to a forward cascade of { odd energy} and an inverse cascade of wave action. The solution corresponding to the forward cascade is non-local (the collision integral diverges), and thus it cannot be observed. The solution for the inverse cascade, however, is only marginally non-local, and is therefore valid up to a logarithmic correction. We were able to observe this inverse cascade, which corresponds once again to a spectrum $N(k)\propto k^{-3}$, in our numerical simulations, which we discuss in the last section. Finally, we also use numerical simulations and phenomenological arguments to explore what happens when the strength of the nonlinearity increases beyond the weak wave turbulence regime.

\subsubsection{Canonical transformation}

In this subsection, we derive the coefficient of the six-wave interaction following the same approach as \cite{Zakharov1992,Krasitskii_1994,Laurie2012,Boffetta2009,Laurie2010}. 
The idea is to perform a change of variable from $A(t,k)$ to a new variable $c(t,k)$. This change of variable should be canonical, i.e. it should conserve the Hamilton equation \eqref{Hamilton_eq_a}. 
For this, we take advantage of the fact that the time evolution operation is canonical, and introduce a third auxiliary variable $b(t',k)$, such that $b(0,k)=c(t,k)$ and $b(1,k)=A(t,k)$. This auxiliary variable evolves according to an auxiliary Hamiltonian $\mathcal{H}_{\rm aux}$, with arbitrary coefficients. We then write the old variable $A_k=b(1,k)$ as a Taylor series around $t'=0$,
\be \label{TaylorCT}
A_k = c_k + \left(\frac{\partial b(t',k)}{\partial t'}\right)\bigg\rvert_{t'=0} + \frac{1}{2} \left(\frac{\partial^2 b(t',k)}{\partial t'^2}\right)\bigg\rvert_{t'=0} +...
\ee
with
\bea \label{CTderivatives_general}
&&\frac{\partial b(t',k)}{\partial t'} = -i\frac{\delta \mathcal{H}_{\rm aux}}{\delta b_k^*} \;, \\
&&\frac{\partial^2 b(t',k)}{\partial t'^2} = -i \frac{\partial}{\partial t'} \frac{\delta \mathcal{H}_{\rm aux}}{\delta b_k^*} \nn \;.
\eea
We now have to write the most general form of the Hamiltonian $\mathcal{H}_{\rm aux}$. For our computation we only need to include terms up to order six (and starting at order three, since we do not want to modify the linear term). In addition, the new Hamiltonian should also conserve the wave action, which strongly reduces the number of terms that we need to include. We can therefore consider a Hamiltonian of the form
\be \label{H_aux}
\mathcal{H}_{\rm aux} = \frac{1}{4} \int_{1,2,3} \delta_{12}^{34} \, \tilde T_{12}^{34} \, b_1^* b_2^* b_3 b_4 + \frac{1}{36} \int_{\{k_i\}} \delta_{123}^{456} \, \tilde W_{123}^{456} \, b_1^* b_2^* b_3^* b_4 b_5 b_6 \;,
\ee
where contrary to \cite{Laurie2012} we have directly set the non-wave action conserving terms to zero. In the above expression, the coefficient $\tilde T$ satisfies the symmetries $\tilde T_{12}^{34}=\tilde T_{21}^{34}=\tilde T_{12}^{43}=\tilde T_{21}^{43}$, and similarly for $\tilde W$. In addition, for the Hamiltonian to be real, these two coefficients should satisfy the relations $(\tilde T_{12}^{34})^*=\tilde T_{34}^{12}$ and $(\tilde W_{123}^{456})^*=\tilde W_{456}^{123}$.

We then have, using \eqref{CTderivatives_general}
\be \label{CTderivative1}
\frac{\partial b(t',k)}{\partial t'} = -\frac{i}{2} \int_{1,2,3} \delta_{k1}^{23} \tilde T_{k1}^{23} \, b_1^* b_2 b_3 - \frac{i}{12} \int_{\{k_i\}} \delta_{k12}^{345} \, \tilde W_{k12}^{345} \, b_1^* b_2^* b_3 b_4 b_5 \;,
\ee
and, omitting higher order terms,
\bea
\frac{\partial^2 b(t',k)}{\partial t'^2} &=& -\frac{i}{2} \frac{\partial}{\partial t'} \int_{1,2,3} \delta_{k1}^{23} \tilde T_{k1}^{23} \, b_1^* b_2 b_3 + ... \\
&=& - \frac{1}{4} \int_{1,2,3,4,5} \delta_{k12}^{345} \left[ 2 \tilde T_{k1}^{3(k+1-3)} \tilde T_{(4+5-2)2}^{45} - \tilde T_{12}^{3(1+2-3)} \tilde T_{(4+5-k)k}^{45} \right] b_1^* b_2^* b_3 b_4 b_5 +... \nn
\eea
Inserting into \eqref{TaylorCT}, we obtain the expression of $A_k$ as a function of the new variable $c_k$,
\be \label{CTexpr_general}
A_k = c_k - \frac{i}{2} \int_{1,2,3} \delta_{k1}^{23} \, \tilde T_{k1}^{23} c_1^* c_2 c_3 - \int_{\{k_i\}}  \delta_{k12}^{345} \, \Big[ \frac{i}{12} \, \tilde W_{k12}^{345} + \frac{1}{8}  \big( 2\tilde T_{k1}^{3(k+1-3)} \tilde T_{(4+5-2)2}^{45} - \tilde T_{12}^{3(1+2-3)} \tilde T_{(4+5-k)k}^{45} \big) \Big] c_1^* c_2^* c_3 c_4 c_5
\ee
(plus higher order terms). Inserting back into the Hamiltonian \eqref{Hamiltonian1}, this gives, after symmetrization
\bea
\mathcal{H} &=& \int_k \omega_k \, c_k^* c_k + \frac{1}{8} \int_{1,2,3,4} \delta_{12}^{34} \, \big[ T_{12}^{34} - i (\omega_1+\omega_2-\omega_3-\omega_4) \, \tilde T_{12}^{34} \big] c_1^* c_2^* c_3 c_4 \\
&& + \frac{1}{8} \int_{1,...,6} \delta_{123}^{456} \, (\omega_{1+2-4} + \omega_4 - \omega_1-\omega_2 + \omega_{5+6-3} + \omega_3 - \omega_5 - \omega_6) \, \tilde T_{12}^{(1+2-4)4} \tilde T_{(5+6-3)3}^{56} \, c_1^* c_2^* c_3^* c_4 c_5 c_6 \nn \\
&& + \frac{i}{4} \int_{1,...,6} \delta_{123}^{456} \left[ T_{(5+6-3)3}^{56} \tilde T_{12}^{(1+2-4)4} - T_{(1+2-4)1}^{24} \tilde T_{(5+6-3)3}^{56} \right] c_1^* c_2^* c_3^* c_4 c_5 c_6 \nn \\
&& - \frac{i}{36} \int_{1,...,6} \delta_{123}^{456} \, (\omega_1+\omega_2+\omega_3-\omega_4-\omega_5-\omega_6) \, \tilde W_{123}^{456} \, c_1^* c_2^* c_3^* c_4 c_5 c_6 \;. \nn
\eea
We now see that the four-wave contributions can be eliminated by imposing
\be \label{Ttilde_expr}
\tilde T_{12}^{34} = \frac{-i \, T_{12}^{34}}{\omega_1+\omega_2-\omega_3-\omega_4} \;.
\ee
The expression \eqref{Ttilde_expr} is well-defined since the denominator only vanishes for trivial interactions, and one can check that it satisfies the required symmetries. With this new variable, the lowest order of interaction in the Hamiltonian will thus be six-wave.

Using the expression for $\tilde T$ given in \eqref{Ttilde_expr}, we finally obtain
\be \label{Hamiltonian_6wave_full}
\mathcal{H} = \int_k \omega_k\,  c_k^* c_k + \frac{1}{36} \int_{1,...,6} \delta_{123}^{456} \, \big[ \mathcal{W}_{123}^{456} - i (\omega_1+\omega_2+\omega_3-\omega_4-\omega_5-\omega_6) \tilde W_{123}^{456} \big] c_1^* c_2^* c_3^* c_4 c_5 c_6 \;,
\ee
with, after symmetrization,
\bea \label{defW}
\mathcal{W}_{123}^{456} &=& - \frac{1}{8} \sum_{\substack{i,j,m=1\\i\neq j \neq m \neq i}}^3 \sum_{\substack{p,q,r=4\\p\neq q \neq r \neq p}}^6 \left[ \frac{1}{\omega_{i+j+p}+\omega_p-\omega_i-\omega_j} + \frac{1}{\omega_{q+r-m}+\omega_m-\omega_q-\omega_r} \right] T_{(i+j-p)p}^{ij} T_{(q+r-m)m}^{qr} \\
&=& - \frac{1}{2} \frac{(G^{\text{NL}}_{\text{odd}})^2}{G_{\rm odd}} k_1k_2k_3k_4k_5k_6  \sum_{\substack{i,j,m=1\\i\neq j \neq m \neq i}}^3 \sum_{\substack{p,q,r=4\\p\neq q \neq r \neq p}}^6 (k_i+k_j-k_p) (k_q+k_r-k_m) \nn \\
&& \hspace{1.5cm} \times \left[ \frac{1}{(k_i+k_j-k_p)^2 + k_p^2 -k_i^2 -k_j^2} + \frac{1}{(k_q+k_r-k_m)^2+k_m^2-k_q^2-k_r^2} \right] \;, \nn
\eea
where in the final step we have used that $\omega_k=G_{\rm odd}k^2$ and $T_{12}^{34}=2\,G^{\text{NL}}_{\text{odd}}\, k_1k_2k_3k_4$. The result in terms of $T_{12}^{34}$ and $\omega_k$ is the same as the one given in \cite{Laurie2012,Boffetta2009,Laurie2010}. The term involving the coefficient $\tilde W_{123}^{456}$ vanishes at the six-wave resonance, and therefore it is irrelevant for our computations. We can thus choose this coefficient to be always zero. This leads to the final expression for the Hamiltonian,
\be \label{Hamiltonian_6wave_final}
\mathcal{H} = \int_k \omega_k\,  c_k^* c_k + \frac{1}{36} \int_{1,...,6} \delta_{123}^{456} \, \mathcal{W}_{123}^{456} \, c_1^* c_2^* c_3^* c_4 c_5 c_6 \;.
\ee

The factor $\mathcal{W}$ is real and has the following symmetries
\be \label{symW}
\mathcal{W}_{123}^{456} = \mathcal{W}_{213}^{456} = \mathcal{W}_{321}^{456} = ... = \mathcal{W}_{123}^{546} = \mathcal{W}_{123}^{654} = ... = \mathcal{W}_{456}^{123} \;,
\ee
and the homogeneity property
\be \label{homW}
\mathcal{W}(\lambda k_1, \lambda k_2, \lambda k_3; \lambda k_4, \lambda k_5, \lambda k_6) = \lambda^6 \mathcal{W}(k_1,k_2,k_3;k_4,k_5,k_6) \;.
\ee

Since the transformation is canonical, we can directly obtain from the Hamiltonian \eqref{Hamiltonian_6wave_final} the evolution equation for the new variable $c(t,k)$,
\be \label{eq6wave}
i\partial_t c_k = \frac{\delta \mathcal{H}}{\delta c_k^*} = \omega_k c_k + \frac{1}{12} \int_{1,2,3,4,5} \delta_{k12}^{345} \, \mathcal{W}_{k12}^{345} \, c_1^* c_2^* c_3 c_4 c_5 \;.
\ee

\subsubsection{Wave kinetic equation}

We now derive the kinetic equation for the wave action spectrum $n(k)$, following { the random phase approximation method as presented in} \cite{Zakharov1992}.  
To leading order we still have 
\be
n(k) \delta(k-k') = \langle c_k c_{k'}^* \rangle \;.
\ee
We start from equation \eqref{eq6wave}. Multiplying this equation by $c_{k'}^*$, subtracting the complex conjugate with $k\leftrightarrow k'$, averaging and multiplying by $-i$ we obtain
\be \label{eq_nk_1}
\delta_k^{k'} \partial_t n_k = -\frac{i}{12} \int_{1,2,3,4,5} \left[ \delta_{k12}^{345} \mathcal{W}_{k12}^{345} \langle c_{k'}^* c_1^* c_2^* c_3 c_4 c_5 \rangle - \delta_{k'12}^{345} \mathcal{W}_{k'12}^{345} \langle c_{k} c_1 c_2 c_3^* c_4^*  c_5^* \rangle \right]  \;, 
\ee
where we denote $\delta_k^{k'}=\delta(k-k')$. To leading order, the correlations of order 6 take the form
\be \label{c_corr6}
\langle c_1 c_2 c_3 c_4^* c_5^* c_6^* \rangle = n_1 n_2 n_3 \sum_{\substack{p,q,r=4\\p\neq q \neq r \neq p}}^6 \delta_1^p \delta_2^q \delta_3^r \;.
\ee
Inserting this into \eqref{eq_nk_1}, we see however that the r.h.s vanishes, and thus we need to go to the next order. For this, we use again equation \eqref{eq6wave} to write,
\begin{flalign}
&(i\partial_t + \omega_1 + \omega_2 + \omega_3 - \omega_4 - \omega_5 - \omega_6) \langle c_1^* c_2^* c_3^* c_4 c_5 c_6 \rangle \\
& = -\frac{1}{12} \int_{a,b,c,d,e} \left[ \mathcal{W}_{1ab}^{cde} \delta_{1ab}^{cde} \langle c_2^* c_3^* c_4 c_5 c_6 c_a c_b c_c^* c_d^* c_e^* \rangle + \mathcal{W}_{2ab}^{cde} \delta_{2ab}^{cde} \langle c_1^* c_3^* c_4 c_5 c_6 c_a c_b c_c^* c_d^* c_e^* \rangle  + \mathcal{W}_{3ab}^{cde} \delta_{3ab}^{cde} \langle c_1^* c_2^* c_4 c_5 c_6 c_a c_b c_c^* c_d^* c_e^* \rangle \right. \nn \\
& \hspace{0.2cm} \left. - \mathcal{W}_{4ab}^{cde} \delta_{4ab}^{cde} \langle c_1^* c_2^* c_3^* c_5 c_6 c_a^* c_b^* c_c c_d c_e \rangle  - \mathcal{W}_{5ab}^{cde} \delta_{5ab}^{cde} \langle c_1^{s_1*} c_2^{s_2*} c_3^{s_3*} c_4^{s_4} c_6^{s_6} c_a^{s_a*} c_b^{s_b*} c_c^{s_c} c_d^{s_d} c_e^{s_e} \rangle - \mathcal{W}_{6ab}^{cde}  \delta_{6ab}^{cde} \langle c_1^* c_2^* c_3^* c_4 c_5 c_a^* c_b^* c_c c_d c_e \rangle \right] \nn \\
& = - \delta_{123}^{456} \, W_{123}^{456} n_1 n_2 n_3 n_4 n_5 n_6 \left[ \frac{1}{n_1} + \frac{1}{n_2} + \frac{1}{n_3} - \frac{1}{n_4} - \frac{1}{n_5} - \frac{1}{n_6} \right] \;. \nn
\end{flalign}
Following Zakharov, we neglect the oscillating part, which varies on a timescale much faster than the non-linear timescale. This leads to
\bea
\langle c_1^* c_2^* c_3^* c_4 c_5 c_6 \rangle &=& - \delta_{123}^{456} \, \mathcal{W}_{123}^{456} \frac{n_1 n_2 n_3 n_4 n_5 n_6 }{\omega_1 + \omega_2 + \omega_3 - \omega_4 - \omega_5 - \omega_6 + i\delta} \left[ \frac{1}{n_1} + \frac{1}{n_2} + \frac{1}{n_3} - \frac{1}{n_4} - \frac{1}{n_5} - \frac{1}{n_6} \right] \nn \\
&:=& \delta_{123}^{456} \, C_{123}^{456} \;.
\eea
Finally, inserting into \eqref{eq_nk_1} and using ${\rm Im}\frac{1}{\omega+i\delta} = -\pi \delta(\omega)$, we arrive at the kinetic equation
\bea \label{kinetic1}
\partial_t n_k &=& \frac{1}{6} \int_{12345} \, \delta_{k12}^{345} \mathcal{W}_{k12}^{345} \, {\rm Im} [ C_{k12}^{345}] \\
&=& \frac{\pi}{6} \int_{12345} \, (\mathcal{W}_{k12}^{345})^2 n_k n_1 n_2 n_3 n_4 n_5 \left[ \frac{1}{n_k} + \frac{1}{n_1} + \frac{1}{n_2} - \frac{1}{n_3} - \frac{1}{n_4} - \frac{1}{n_5} \right] \delta_{k12}^{345} \, \delta(\omega_k + \omega_1 + \omega_2 - \omega_3 - \omega_4 - \omega_5) \;. \nn
\eea
\\

\subsubsection{Detailed conservation laws}

Introducing the quantity
\be
\Phi = \int dk \rho_k n_k \;,
\ee
and using the kinetic equation \eqref{kinetic1}, we can write after symmetrization
\bea
\partial_t \Phi &=& \int dk \rho_k \partial_t n_k \\
&=& \frac{\pi}{36} \int_{k12345} \, (\mathcal{W}_{k12}^{345})^2 n_k n_1 n_2 n_3 n_4 n_5 \left[ \frac{1}{n_k} + \frac{1}{n_1} + \frac{1}{n_2} - \frac{1}{n_3} - \frac{1}{n_4} - \frac{1}{n_5} \right] \left[\rho_k + \rho_1 + \rho_2 - \rho_3 - \rho_4 - \rho_5 \right]  \nn \\
&& \hspace{1.6cm} \times \delta(k+k_1+k_2-k_3-k_4-k_5) \delta(\omega_k +  \omega_1 + \omega_2 - \omega_3 - \omega_4 - \omega_5) \;. \nn
\eea
From this expression we immediately see that the following quantities are locally conserved:

\noindent (i) the wave action density, with $\rho_k=1$: $\bar N = \int dk \, n_k$,

\noindent (ii) the odd energy density, with $\rho_k=\omega_k$: $\bar E = \int dk \, \omega_k n_k$,

\noindent (iii) the momentum density, with $\rho_k=k$: $\bar P=\int dk \, k n_k$.

{These three quantities correspond to the densities in physical space of the three quantities that are conserved at the scale of the full system, exactly for the wave action and momentum, and approximately in the weakly non-linear limit for the { odd energy}. This means that, on average, in the weak wave turbulence regime, these three quantities are conserved for any subpart of the system, as expected from the statistically homogeneous turbulence hypothesis.}

\subsubsection{Zakharov-Kolmogorov spectra}

We now look for stationary solutions of the kinetic equation \eqref{kinetic1} using the Kuznetsov-Zakharov transformation \cite{Zakharov1992}. From here on, we assume that the spectrum is even in $k$, $n(k)=n(-k)$. This implies that $N(k)=n(k)$ in \eqref{defN}, and that the odd momentum $P$ vanishes.

Let us introduce, for $k\in\mathbb{R}^+$, the function $f(k)$ such that
\be
f(k) = \omega_k = |G_{\rm odd}| k^2 \;.
\ee
For $k\geq0$ this function is invertible, i.e. one can write $k=f^{-1}(\omega_k)=\sqrt{\omega_k/|G_{\rm odd}|}$. We can thus define
\be
\tilde N(\omega) = N(f^{-1}(\omega)) \;.
\ee
Let us start by rewriting the kinetic equation \eqref{kinetic1} using only integrals over positive $k$'s. For $k\geq0$ we can write
\bea \label{kinetic_pos}
\partial_t N_k &=& \sum_{\{\sigma_i=\pm 1\}} \int_0^\infty dk_1dk_2dk_3dk_4dk_5 \, \mathcal{W}_{k12}^{345} \, N_k N_1 N_2 N_3 N_4 N_5 \left[ \frac{1}{N_k} + \frac{1}{N_1} + \frac{1}{N_2} - \frac{1}{N_3} - \frac{1}{N_4} - \frac{1}{N_5} \right] \\
&&\hspace{1cm} \times \delta(k+\sigma_1 k_1+\sigma_2 k_2-\sigma_3 k_3-\sigma_4 k_4-\sigma_5 k_5) \delta(\omega_k + \omega_1 + \omega_2 - \omega_3 - \omega_4 - \omega_5) \;. \nn 
\eea
We then perform a change of variables
\be
\omega_1 = f(k_1) \quad \Rightarrow \quad dk_1 = \frac{1}{f'(f^{-1}(\omega_1))} d\omega_1 = \frac{d\omega_1}{2\sqrt{|G_{\rm odd}| \omega_1}} \;,
\ee
and similarly for $k_2$ and $k_3$. Introducing
\be \label{relN_omega}
\mathcal{\tilde N}(\omega) = \frac{1}{2\sqrt{|G_{\rm odd}| \omega}} \tilde N(\omega) \;,
\ee
the kinetic equation can be rewritten as
\be
\frac{\partial\mathcal{\tilde N}(\omega,t)}{\partial t} = \mathcal{I}(\omega) \;,
\ee
with
\bea \label{kinetic_omega}
\mathcal{I}(\omega) &=& \int_0^\infty d\omega_1 d\omega_2 d\omega_3 d\omega_4 d\omega_5 \, U(\omega,\omega_1,\omega_2;\omega_3,\omega_4,\omega_5) \, \tilde N(\omega) \tilde N(\omega_1) \tilde N(\omega_2) \tilde N(\omega_3) \tilde N(\omega_4) \tilde N(\omega_5)  \\
&&\hspace{0.5cm} \times \left[ \frac{1}{\tilde N(\omega)} + \frac{1}{\tilde N(\omega_1)} + \frac{1}{\tilde N(\omega_2)} - \frac{1}{\tilde N(\omega_3)} - \frac{1}{\tilde N(\omega_4)} - \frac{1}{\tilde N(\omega_5)} \right] \delta(\omega + \omega_1 + \omega_2 - \omega_3 - \omega_4 - \omega_5) \;, \nn 
\eea
where
\bea \label{defU}
U(\omega,\omega_1,\omega_2;\omega_3,\omega_4,\omega_5) &=& \frac{1}{64} \frac{\mathcal{W}_{k12}^{345}}{|G_{\rm odd}|^3 \sqrt{\omega\omega_1\omega_2\omega_3\omega_4\omega_5}} \\
&& \times \sum_{\{\sigma_i=\pm 1\}} \delta(k+\sigma_1 k_1+\sigma_2 k_2-\sigma_3 k_3-\sigma_4 k_4-\sigma_5 k_5)|_{\{k_i=\sqrt{\frac{\omega_i}{|G_{\rm odd}|}}\}} \;. \nn
\eea
The function $U$ has the same symmetry properties as $\mathcal{W}_{k12}^{345}$ [see \eqref{symW}],
\bea \label{Usym}
U(\omega,\omega_1,\omega_2;\omega_3,\omega_4,\omega_5) &=& U(\omega_1,\omega,\omega_2;\omega_3,\omega_4,\omega_5) = U(\omega_2,\omega_1,\omega;\omega_3,\omega_4,\omega_5) = ... \\
&=& U(\omega,\omega_1,\omega_2;\omega_4,\omega_3,\omega_5) = U(\omega,\omega_1,\omega_2;\omega_5,\omega_4,\omega_3) = ... \nn \\
&=& U(\omega_3,\omega_4,\omega_5;\omega,\omega_1,\omega_2) \;,
\eea
as well as the homogeneity property [see \eqref{homW}]
\be \label{U_factorization}
U(\lambda\omega,\lambda\omega_1,\lambda\omega_2;\lambda\omega_3,\lambda\omega_4,\lambda\omega_5) = \lambda^{5/2} \, U(\omega,\omega_1,\omega_2;\omega_3,\omega_4,\omega_5)  \;.
\ee

Let us now assume that the wave action spectrum takes a power law form, i.e. $\tilde N(\omega)=A\omega^{-x}$. 
Equation \eqref{kinetic_omega} becomes
\bea \label{kinetic_ansatz}
\mathcal{I}(\omega) &=& A^5 \int_0^\infty d\omega_1 d\omega_2 d\omega_3 d\omega_4 d\omega_5 \, U(\omega,\omega_1,\omega_2;\omega_3,\omega_4,\omega_5)[\omega\omega_1\omega_2\omega_3\omega_4\omega_5]^{-x} \\
&& \hspace{1.2cm} \times [ \omega^x + \omega_1^x + \omega_2^x - \omega_3^x - \omega_4^x - \omega_5^x ] \delta(\omega + \omega_1 + \omega_2 - \omega_3 - \omega_4 - \omega_5) \;. \nn
\eea
The integration domain can be divided into 6 regions,
\bea
&&\Omega_0=\{\omega_1,\omega_2,\omega_3,\omega_4,\omega_5<\omega\} \;, \\
&&\Omega_i=\{\omega_i>\omega, \ \omega_i>\omega_j \ \forall j\neq i\} \quad \text{for } i=1,...,5 \;. \nn 
\eea
The idea of the KZ transformation is to apply a conformal transformation to map each region onto $\Omega_0$. Let us denote $I_i(\omega)$ the integral over the region $\Omega_i$. For the region $\Omega_1$ we write
\be
\omega_1 = \frac{\omega^2}{\omega_1'} \quad , \quad \omega_2 = \frac{\omega \omega_2'}{\omega_1'} \quad , \quad \omega_3 = \frac{\omega \omega_3'}{\omega_1'} \quad , \quad \omega_4 = \frac{\omega \omega_4'}{\omega_1'} \quad , \quad \omega_5 = \frac{\omega \omega_5'}{\omega_1'} \;.
\ee
One can check that this indeed maps the domain $\Omega_1$ onto $\Omega_0$. The Jacobian of this transformation is $\frac{\omega^6}{\omega_1'^6}$. Using the factorization property \eqref{U_factorization} with $\lambda=\frac{\omega}{\omega_1'}$, we obtain (dropping the prime)
\bea \label{kinetic_region1}
\mathcal{I}_1(\omega) &=& A^5 \int_{\Omega_0} d\omega_1 d\omega_2 d\omega_3 d\omega_4 d\omega_5 \, \left(\frac{\omega}{\omega_1}\right)^{\frac{15}{2}-5x} U(\omega_1,\omega,\omega_2;\omega_3,\omega_4,\omega_5) [\omega\omega_1\omega_2\omega_3\omega_4\omega_5]^{-x} \\
&& \hspace{1.2cm} \times [ \omega_1^x + \omega^x + \omega_2^x - \omega_3^x - \omega_4^x - \omega_5^x ] \delta(\omega_1 + \omega + \omega_2 - \omega_3 - \omega_4 - \omega_5) \;. \nn
\eea
For the region $\Omega_2$, a similar transformation (with $\omega_1$ and $\omega_2$ exchanged) leads to
\bea \label{kinetic_region2}
\mathcal{I}_2(\omega) &=& A^5 \int_{\Omega_0} d\omega_1 d\omega_2 d\omega_3 d\omega_4 d\omega_5 \, \left(\frac{\omega}{\omega_2}\right)^{\frac{15}{2}-5x} U(\omega_2,\omega_1,\omega;\omega_3,\omega_4,\omega_5) [\omega\omega_1\omega_2\omega_3\omega_4\omega_5]^{-x} \\
&& \hspace{1.2cm} \times [ \omega^x + \omega_1^x + \omega_2^x - \omega_3^x - \omega_4^x - \omega_5^x ] \delta(\omega + \omega_1 + \omega_2 - \omega_3 - \omega_4 - \omega_5) \;. \nn
\eea
For the region $\Omega_3$, the transformation reads
\be
\omega_1 = \frac{\omega\omega_4'}{\omega_3'} \quad , \quad \omega_2 = \frac{\omega\omega_5'}{\omega_3'} \quad , \quad \omega_3 = \frac{\omega^2}{\omega_3'} \quad , \quad \omega_4 = \frac{\omega\omega_1'}{\omega_3'} \quad , \quad \omega_5 = \frac{\omega\omega_2'}{\omega_3'} \;.
\ee
One can again check that this maps the domain $\Omega_3$ onto $\Omega_0$. The Jacobian of this transformation is $\frac{\omega^6}{\omega_3'^6}$. Using the factorization property \eqref{U_factorization} with $\lambda=\frac{\omega}{\omega_3'}$, we obtain (dropping the prime)
\bea \label{kinetic_region3}
\mathcal{I}_3(\omega) &=& -A^5 \int_{\Omega_0} d\omega_1 d\omega_2 d\omega_3 d\omega_4 d\omega_5 \, \left(\frac{\omega}{\omega_3}\right)^{\frac{15}{2}-5x} U(\omega_3,\omega_4,\omega_5;\omega,\omega_1,\omega_2) [\omega\omega_1\omega_2\omega_3\omega_4\omega_5]^{-x} \\
&& \hspace{1.2cm} \times [ \omega^x + \omega_1^x + \omega_2^x - \omega_3^x - \omega_4^x - \omega_5^x ] \delta(\omega + \omega_1 + \omega_2 - \omega_3 - \omega_4 - \omega_5) \;. \nn
\eea
Finally, performing similar transformations for the last two regions $\Omega_4$ and $\Omega_5$ leads to
\bea \label{kinetic_region4}
\mathcal{I}_4(\omega) &=& -A^5 \int_{\Omega_0} d\omega_1 d\omega_2 d\omega_3 d\omega_4 d\omega_5 \, \left(\frac{\omega}{\omega_4}\right)^{\frac{15}{2}-5x} U(\omega_4,\omega_5,\omega_3;\omega_2,\omega,\omega_1) [\omega\omega_1\omega_2\omega_3\omega_4\omega_5]^{-x} \\
&& \hspace{1.2cm} \times [ \omega^x + \omega_1^x + \omega_2^x - \omega_3^x - \omega_4^x - \omega_5^x ] \delta(\omega + \omega_1 + \omega_2 - \omega_3 - \omega_4 - \omega_5) \;. \nn
\eea
and
\bea \label{kinetic_region5}
\mathcal{I}_5(\omega) &=& -A^5 \int_{\Omega_0} d\omega_1 d\omega_2 d\omega_3 d\omega_4 d\omega_5 \, \left(\frac{\omega}{\omega_5}\right)^{\frac{15}{2}-5x} U(\omega_5,\omega_3,\omega_4;\omega_1,\omega_2,\omega) [\omega\omega_1\omega_2\omega_3\omega_4\omega_5]^{-x} \\
&& \hspace{1.2cm} \times [ \omega^x + \omega_1^x + \omega_2^x - \omega_3^x - \omega_4^x - \omega_5^x ] \delta(\omega + \omega_1 + \omega_2 - \omega_3 - \omega_4 - \omega_5) \;. \nn
\eea

Summing the 6 integrals and using the symmetry properties \eqref{Usym} to factorize the function $U$, we finally obtain
\bea \label{kinetic_Zakharov_final}
\mathcal{I}(\omega) &=& A^5 \int_{\Omega_0} d\omega_1 d\omega_2 d\omega_3 d\omega_4 d\omega_5 \, U(\omega,\omega_1,\omega_2;\omega_3,\omega_4,\omega_5) [\omega\omega_1\omega_2\omega_3\omega_4\omega_5]^{-x} \\
&& \hspace{1cm} \times [ \omega^x + \omega_1^x + \omega_2^x - \omega_3^x - \omega_4^x - \omega_5^x ] \left[ 1 + \left(\frac{\omega_1}{\omega}\right)^{y} + \left(\frac{\omega_2}{\omega}\right)^{y} - \left(\frac{\omega_3}{\omega}\right)^{y} - \left(\frac{\omega_4}{\omega}\right)^{y} - \left(\frac{\omega_5}{\omega}\right)^{y} \right] \nn \\
&& \hspace{1cm} \times \delta(\omega + \omega_1 + \omega_2 - \omega_3 - \omega_4 - \omega_5) \;, \nn
\eea
with
\be
y=5x-\frac{15}{2} \;.
\ee
From equation \eqref{kinetic_Zakharov_final}, one can directly read the stationary solutions. First, there are two zero-flux solutions given by $x=0$ and $x=1$ (i.e. $N(k)\propto|k|^0$ and $N(k)\propto|k|^{-2}$). These two solutions, sometimes called thermodynamic solutions, correspond respectively to the equipartition of wave action and equipartition of odd energy. We will not study them here. In addition, we have two non-equilibrium solutions which are given by
\bea \label{sol_x}
&& y = 0 \quad \Rightarrow \quad x = \frac{3}{2} \;, \\
&& y = 1 \quad \Rightarrow \quad x = \frac{17}{10} \;.
\eea
In terms of $k$, this reads
\be \label{sol_k}
N(k)\propto |k|^{-3} \quad \text{or} \quad N(k)\propto |k|^{-17/5} = |k|^{-3.4} \;.
\ee
These two non-equilibrium solutions are the Kolmogorov-Zakharov spectra. Each of these solution corresponds to a cascade of one of the two conserved quantities, namely the wave action or the { odd energy}, in the inertial range which separates the forcing scale and the dissipation scale. In the rest of this section, we will study the different properties of these two solutions, in particular the nature and the direction of the associated cascades. We also need to make sure that they are valid solutions of the kinetic equation by checking the convergence of the collision integral in \eqref{kinetic1}.

\subsubsection{Nature of the cascades}

Let us now consider the kinetic equation \eqref{kinetic1} with $n(k)=N(k)=C|k|^{-\nu}$ and rescale the integral to take out all the $k$ dependence by writing $k_i=k z_i$ for $i=1,2,3,4,5$. This gives
\be \label{kinetic_rescaled}
\partial_t N_k = \frac{(G^{\text{NL}}_{\text{odd}})^4}{|G_{\rm odd}|^{3}} C^5 k^{14-5\nu} I(\nu) \;,
\ee
where
\bea \label{Int1}
&& \hspace{-0.5cm} I(\nu) = \frac{\pi}{6} \int dz_1dz_2dz_3dz_4dz_5 \, (w_{1z_1z_2}^{z_3z_4z_5})^2 (z_1z_2z_3z_4z_5)^{-\nu} \left[ 1+z_1^\nu+z_2^\nu-z_3^\nu-z_4^\nu-z_5^\nu \right] \nn \\
&&\hspace{1.5cm} \times \delta(1+z_1+z_2-z_3-z_4-z_5) \delta(1 + z_1^2 + z_2^2 - z_3^2 - z_4^2 - z_5^2) \;,
\eea
and
\bea \label{defWplus_adim}
w_{1z_1z_2}^{z_3z_4z_5} &=& - z_1z_2z_3z_4z_5z_6  \sum_{\substack{i,j,m=1\\i\neq j \neq m \neq i}}^3 \sum_{\substack{p,q,r=4\\p\neq q \neq r \neq p}}^6 \left[ \frac{(z_p+z_q-z_i) (z_j+z_m-z_r)}{(z_j+z_m-z_r)^2 +z_r^2 -z_j^2 - z_m^2} \right. \nn \\
&& \hspace{3.5cm} \left. + \frac{(z_i+z_j-z_p) (z_q+z_r-z_m)}{(z_q+z_r-z_m)^2+z_m^2-z_q^2-z_r^2} \right] \;.
\eea
Since the integral $I(\nu)$ is proportional to the integral $\mathcal{I}(\omega)$ of Eq.~\eqref{kinetic_Zakharov_final}, we know that it vanishes for the values of $\nu$ corresponding to the thermodynamic and KZ solutions. We can rewrite the equation \eqref{kinetic_rescaled} in the form of a continuity equation for the wave action,
\be \label{continuity_eq}
\partial_t N_k + \partial_k \Pi^N = 0 
\ee
with
\be \label{waveaction_flux}
\Pi^N(k) = -\frac{(G^{\text{NL}}_{\text{odd}})^4}{|G_{\rm odd}|^{3}} \frac{C^5 I(\nu)}{15-5\nu} \, k^{15-5\nu} \;.
\ee
We now see that the solution $\nu=3$ corresponds to a constant wave action flux. Thus, this solution describes the cascade of wave action from the injection scale to the dissipation scale. Below we will show that this is in fact an inverse cascade (from large scales to small scales).

This argument also gives us the scaling of the constant $C$ as a function of the wave action flux and of the parameters of the model. Putting everything together we obtain
\be \label{np_full_spectrum}
N(k) = \mathcal{C}_{\rm N} \, |G_{\rm odd}|^{{3}/5} |G^{\text{NL}}_{\text{odd}}|^{-4/5} \varepsilon_N^{1/5} k^{-3} 
\ee
where $\varepsilon_N=|\Pi^N|$ is the wave action injection rate, and $\mathcal{C}_{\rm N}$ is the Kolmogorov constant (which is dimensionless).

Similarly, one can write a continuity equation for the { odd energy},
\be \label{continuity_eq_energy}
\partial_t (\omega_k N_k) + \partial_k \Pi^E = 0 
\ee
with
\be \label{energy_flux}
\Pi^E(k) = -\frac{(G^{\text{NL}}_{\text{odd}})^4}{G_{\rm odd}^2} \frac{C^5 I(\nu)}{17-5\nu} \, k^{17-5\nu} \;.
\ee
Thus, the solution $\nu=17/5$ corresponds to a constant { odd energy} flux, i.e. it describes the cascade of { odd energy}. It reads
\be \label{energy_full_spectrum}
N(k) = \mathcal{C}_{\rm E} \, {|G_{\rm odd}|^{2/5}} |G^{\text{NL}}_{\text{odd}}|^{-4/5} \varepsilon_E^{1/5} k^{-17/5}
\ee
where $\varepsilon_E=|\Pi^E|$ is the odd energy injection rate, and $\mathcal{C}_{\rm E}$ is another Kolmogorov constant.

\subsubsection{Locality of the interaction}

For the above solutions to be valid, the collision integral in the kinetic wave equation \eqref{kinetic1} needs to converge. This is called the locality hypothesis. If the integral diverges for a given KZ solution, then this solution cannot be realized.

For this section it is convenient to integrate over the two delta functions, leading to the parametrization
\bea
\tilde k_2 &=& k_4 - \frac{(k-k_3) (k_1-k_3)}{k+k_1-k_3-k_4} \;, \\
\tilde k_5 &=& k + k_1 - k_3 - \frac{(k-k_3) (k_1-k_3)}{k+k_1-k_3-k_4} \;. \nn
\eea
Let us start by studying the IR convergence, by taking one of the wavenumbers to zero, e.g. $k_4\to 0$. In this limit, we find from the expression \eqref{defW} that the factor $\mathcal{W}_{123}^{456}$ behaves as
\be
\mathcal{W}_{123}^{456} \sim k_4 \;.
\ee
Taking $N(k)\propto k^{-\nu}$ with $\nu>0$, this leads to
\bea
(\mathcal{W}_{123}^{456})^2 N_k N_1 N_2 N_3 N_4 N_5 \left[ \frac{1}{N_k} + \frac{1}{N_1} + \frac{1}{N_2} - \frac{1}{N_3} - \frac{1}{N_4} - \frac{1}{N_5} \right] \sim k_4^{2-\nu} \;.
\eea
Thus, the integral converges in the IR if
\be
\nu<3 \;.
\ee 

Let us now consider the UV convergence by taking $k_4\to +\infty$. Note that in this limit one has $\tilde k_2 =  k_4 + \mathcal{O}(k_4^{-1})$ (and $\tilde k_5=\mathcal{O}(k_4^0)$). We then find that
\be
\mathcal{W}_{123}^{456} \sim k_4^2 \;.
\ee
Taking into account that as $k_4\to+\infty$, one has $k_2-k_4\propto k_4^{-1}$, leading to $k_2^\nu-k_4^\nu\propto k_4^{\nu-2}$, and taking into account the Jacobian arising from the integration over the delta function involving the $\omega$'s, which gives a factor
\be
|\partial_{k_2} (\omega_k+\omega_1+\omega_2-\omega_3-\omega_4-\omega_{k+k_1+k_2-k_3-k_4})|^{-1} = \frac{1}{2|k+k_1-k_3-k_4|} \;,
\ee
we obtain that the argument of the integral behaves as
\bea
\sim k_4^{3-2\nu} (k_4^0 + k_4^{\nu-2}) \;.
\eea
Thus, the integral converges in the UV if
\be
\nu>2 \;.
\ee 
Putting everything together, we see that the integral is convergent for
\be
2 < \nu < 3 \;.
\ee
The solution corresponding to the { odd energy} cascade ($\nu=3.4$) thus does not satisfy the locality hypothesis and is invalid. However, the other KZ solution, which describes the wave action cascade ($\nu=3$), 
only diverges logarithmically. In such cases, the solution may still be valid up to a logarithmic correction, as we now discuss.

\subsubsection{Logarithmic correction}

Following the argument of \cite{Laurie2012} (first developed by Kraichnan for 2D eddy turbulence \cite{Kraichnan_1971}), we assume that the spectrum for the wave action cascade can be corrected as
\be
N(k) = \mathcal{C}_{\rm N} \, |G_{\rm odd}|^{{3}/5} |G^{\text{NL}}_{\text{odd}}|^{-4/5} \varepsilon_N^{1/5} k^{-3} \ln^{-y}(k\ell_d) \;,
\ee
where $\ell_d$ is the scale at which large scale dissipation occurs. 
We write
\be
\Pi^N(k) = \int^k dk \partial_t N_k \propto \int^k dk \, k^{14} N_k^5 \propto \int^k dk \, k^{-1} \ln^{-5y}(k\ell_d) \;.
\ee
This integrates to some power of $\ln(k\ell_d)$, except for $y=1/5$ for which we obtain $\Pi^N(k)\propto\ln\ln(k\ell_d)$. The argument is that this function varies slowly enough so that we can interpret it as a constant over the inertial range (at least for $k\gg \ell_d^{-1}$), and thus we consider the spectrum
\be \label{log_corrected_spectrum}
N(k) = \mathcal{C}_{\rm N} \, |G_{\rm odd}|^{{3}/5} |G^{\text{NL}}_{\text{odd}}|^{-4/5} \varepsilon_N^{1/5} k^{-3} \ln^{-1/5}(k\ell_d)
\ee
as a constant wave action flux solution of the kinetic equation (for which the collision integral is now convergent thanks to the logarithmic correction). Note that in both \cite{Laurie2012} and \cite{Kraichnan_1971}, the divergence is also in the IR, but the solution corresponds to a forward cascade. We nevertheless expect this argument to remain valid in our case, although we will now see that this solution describes an inverse cascade.

\subsubsection{DAM approximation and direction of the cascades}

Let us now finally investigate the direction of the cascade. We recall that the integral $I(\nu)$ vanishes for $\nu=3$, since this is a stationary solution of the kinetic equation. However the denominator in \eqref{waveaction_flux} also vanishes at this value. Thus, in this case the wave action flux reads instead (omitting the logarithmic correction)
\be \label{waveaction_flux_sol}
\Pi^N(k) = \frac{1}{5} \frac{(G^{\text{NL}}_{\text{odd}})^4}{|G_{\rm odd}|} C^5 I'(3)  \;.
\ee
The sign of the flux is thus given by the sign of $I'(3)$ (since $C>0$ by definition). This is however not easy to determine, even numerically, since we need to evaluate an integral over $\mathbb{R}^3$ which has several singularities. The role of the logarithmic correction in this computation is also unclear.

Another approach is to use the Differential Approximation Model (DAM), which assumes that the interactions are strongly local in order to simplify the kinetic equation \eqref{kinetic1} (see e.g. \cite{Dyachenko1992,Zakharov1999,Laurie2012,Nazarenko11,GaltierCUP2023}). This approximation is only accurate when the locality hypothesis is verified, i.e. for $\nu\in(2,3)$. We start from the kinetic equation \eqref{kinetic_omega} written in terms of $\omega_k=|G_{\rm odd}| k^2$, with $\tilde N(\omega) = N(\sqrt{\omega/|G_{\rm odd}|})$. We first integrate this equation over some arbitrary function $h(\omega)$ and use the symmetries of the integral to write
\bea \label{kinetic_helicity_h}
\int \frac{d\omega \, h(\omega)}{2\sqrt{|G_{\rm odd}| \omega}} \partial_t \tilde N(\omega)  &=& \frac{1}{6} \int_0^\infty d\omega d\omega_1 d\omega_2 d\omega_3 d\omega_4 d\omega_5 \, U(\omega,\omega_1,\omega_2;\omega_3,\omega_4,\omega_5) \, \tilde N(\omega) \tilde N(\omega_1) \tilde N(\omega_2) \tilde N(\omega_3) \tilde N(\omega_4) \tilde N(\omega_5) \nn  \\
&&\hspace{0.5cm} \times \left[ \frac{1}{\tilde N(\omega)} + \frac{1}{\tilde N(\omega_1)} + \frac{1}{\tilde N(\omega_2)} - \frac{1}{\tilde N(\omega_3)} - \frac{1}{\tilde N(\omega_4)} - \frac{1}{\tilde N(\omega_5)} \right] \\
&&\hspace{0.5cm} \times \left[ h(\omega) + h(\omega_1) + h(\omega_2) - h(\omega_3) - h(\omega_4) - h(\omega_5) \right] \delta(\omega + \omega_1 + \omega_2 - \omega_3 - \omega_4 - \omega_5) \;. \nn 
\eea
The idea of the DAM is to assume that the $\omega_i$ only contribute to the integral when they are close to $\omega$, i.e. $\omega_i=\omega(1+p_i)$ with $|p_i|\ll 1$. This allows us to perform a Taylor expansion up to order 2 in $p_i$ to obtain
\be \label{kinetic_helicity_Taylor}
\int d\omega \, \frac{h(\omega)}{2\sqrt{|G_{\rm odd}|\omega}} \partial_t \tilde N(\omega)  = S_0 \int_0^\infty d\omega \, \omega^{\frac{21}{2}} \tilde N(\omega)^6 \frac{\partial^2}{\partial\omega^2} \left( \frac{1}{\tilde N(\omega)} \right) \frac{\partial^2 h(\omega)}{\partial\omega^2} \;,
\ee
with
\bea \label{defS0}
S_0 &=& \frac{1}{24} \int_{-\infty}^{\infty} dp_1 dp_2 dp_3 dp_4 dp_5 \, U(1,1+p_1,1+p_2;1+p_3,1+p_4,1+p_5) (p_1^2+p_2^2-p_3^2-p_4^2-p_5^2)^2 \\
&&\hspace{5cm} \times\delta(p_1+p_2-p_3-p_4-p_5) \;. \nn
\eea
Note that the constant $S_0$ is always positive. An integration by parts finally gives
\be \label{kinetic_helicity_Taylor2}
\int d\omega \, \frac{h(\omega)}{2\sqrt{|G_{\rm odd}|\omega}} \partial_t \tilde N(\omega)  = S_0 \int_0^\infty d\omega \, \frac{\partial^2}{\partial\omega^2} \left[ \omega^{\frac{21}{2}} \tilde N(\omega)^6 \frac{\partial^2}{\partial\omega^2} \left( \frac{1}{\tilde N(\omega)} \right) \right] h(\omega) \;.
\ee
Since this equality should be satisfied for any function $h(\omega)$, we arrive at the equation of the DAM
\be \label{DAM}
 \partial_t \tilde N(\omega) =  2S_0 \sqrt{|G_{\rm odd}|\omega} \,\frac{\partial^2}{\partial\omega^2} \left[ \omega^{\frac{21}{2}} \tilde N(\omega)^6 \frac{\partial^2}{\partial\omega^2} \left( \frac{1}{\tilde N(\omega)} \right) \right] \;.
\ee

The thermodynamic solutions are the solutions for which the quantity
\be \label{defR}
R(\omega) = S_0 \, \omega^{\frac{21}{2}} \tilde N(\omega)^6 \frac{\partial^2}{\partial\omega^2} \left( \frac{1}{\tilde N(\omega)} \right)
\ee
is zero for all values of $\omega$. Assuming a power law form $\tilde N(\omega)=A\omega^{-x}=A\omega^{-\frac{\nu}{2}}$, we obtain
\be \label{exprR}
R(\omega) = S_0 A x(x-1) \omega^{\frac{17}{2}-5x} \;.
\ee
The thermodynamic solutions thus correspond to $x=0$ and $x=1$ (i.e. $\nu=0$ and 2), as predicted using the exact kinetic equation. More generally, if $\tilde N(\omega)$ is a power law we get
\be
 \partial_t \tilde N(\omega) = 2S_0 \sqrt{|G_{\rm odd}|} \, A^5 x(x-1)\left(\frac{17}{2}-5x\right)\left(\frac{15}{2}-5x\right) \omega^{7-5x} \;.
\ee
We thus also recover the two non-equilibrium solutions $x=3/2$ and $x=17/10$ (i.e. $\nu=3$ and $3.4$).

The equation \eqref{DAM} can be written under the form
\be
\partial_t \mathcal{\tilde N}(\omega) + \frac{\partial \tilde \Pi^N(\omega)}{\partial\omega} = 0 \;,
\ee
where we have introduced the wave action flux
\be
\tilde \Pi^N(\omega) = -\frac{\partial R(\omega)}{\partial\omega} \;,
\ee
and where we recall the quantity
\be \label{relN_omega2}
\mathcal{\tilde N}(\omega) = \frac{1}{2\sqrt{|G_{\rm odd}| \omega}} \tilde N(\omega) \;.
\ee
Multiplying by $\frac{d\omega}{dk}$ we recover the continuity equation in the $k$ variable \eqref{continuity_eq} with $\tilde \Pi^N(\omega) = \Pi^N(\sqrt{\omega/|G_{\rm odd}|})$. One can also write a continuity equation for the { odd energy} as
\be
\partial_t [\omega \mathcal{\tilde N}(\omega)] + \frac{\partial \tilde \Pi^E(\omega)}{\partial\omega} = 0 \;,
\ee
where
\be
\tilde \Pi^E(\omega) = -\int_0^\omega d\omega \, \omega \frac{\partial^2 R(\omega)}{\partial\omega^2} = R(\omega) - \omega \frac{R(\omega)}{\partial\omega} \;.
\ee
Assuming again a power law form $\tilde N(\omega)=A\omega^{-x}=A\omega^{-\frac{\nu}{2}}$ leads to
\bea
\tilde \Pi^N(\omega) &=& S_0 A^5 x(x-1)\left(5x-\frac{17}{2}\right) \omega^{\frac{15}{2}-5x} \;, \\
\tilde \Pi^E(\omega) &=& S_0 A^5 x(x-1)\left(5x-\frac{15}{2}\right) \omega^{\frac{17}{2}-5x} \;.
\eea
At the valid solution $x=3/2$, we thus have (up to the logarithmic corrections)
\be
\tilde \Pi^N(\omega) = -\frac{3}{4} S_0 A^5 <0  \quad \text{and} \quad 
\tilde \Pi^E(\omega) = 0 \;,
\ee
where we recall that $A$ and $S_0$ are both positive constants. As we had already anticipated above, this solution thus corresponds to an inverse cascade of wave action.

\subsubsection{Beyond the weak wave regime}

Let us now estimate the domain of validity of our weak wave turbulence prediction. From the kinetic equation \eqref{kinetic1}, we can evaluate the transfer time of the wave action cascade,
\be \label{tau_transfer6}
\tau_{\rm tr} \sim \left(\frac{(G_{\rm odd}^{\rm NL})^4}{G_{\rm odd}^3} k^{14} N(k)^4 \right)^{-1} \;.
\ee
Note that the wave action flux can be estimated from this transfer time as
\be
\Pi^N \sim \frac{kN(k)}{\tau_{\rm tr}} \sim \frac{(G_{\rm odd}^{\rm NL})^4}{G_{\rm odd}^3} k^{15} N(k)^5 \;,
\ee
which is compatible with \eqref{np_full_spectrum}. One way to estimate the range of validity of the weak wave turbulence theory is to compare this transfer time to the linear time scale, which gives the ratio
\be \label{tau_transferVStau_lin}
\frac{\tau_{\rm lin}}{\tau_{\rm tr}} \sim \left(\frac{G_{\rm odd}^{\rm NL}}{G_{\rm odd}} \right)^4 k^{12} N(k)^4  \;.
\ee
The weak wave turbulence regime is valid when this ratio is much smaller than 1. For $N(k)\propto k^{-3}$, this ratio is independent of $k$, which means that we can expect the KZ solution corresponding to the inverse cascade to be valid at all scales as long as $G_{\rm odd}^{\rm NL} \ll G_{\rm odd}$.

Another equivalent way to check whether we are in the weak regime is to compute the time scale ratio $\chi \equiv \tau_\textrm{lin}/\tau_\textrm{NL}$, which compares the linear time scale $\tau_\textrm{lin}=(G_\textrm{odd}k^2)^{-1}$ and the non-linear time scale $\tau_\textrm{NL}=(G_\textrm{odd}^\textrm{NL} k^5 N(k))^{-1}$ estimated from the equation of motion \eqref{eom_incomp_soft_solid_overdamped},
\begin{equation}\label{eq:chi_elastic}
    \chi = \frac{\tau_\textrm{lin}}{\tau_\textrm{NL}} \sim \frac{G_\textrm{odd}^\textrm{NL}}{G_\textrm{odd}} k^3 N(k) \;.
\end{equation}
If $\chi \ll 1$, then we are in the weak regime. When $\chi$ is of order one, we enter the critical balance regime. This corresponds to a wave action spectrum
\begin{equation}\label{eq:spectrum_cb_elastic}
    N(k) \sim G_\textrm{odd} (G_\textrm{odd}^\textrm{NL})^{-1} k^{-3}.
\end{equation}
The fact that this spectrum has the same power law $k^{-3}$ as the KZ solution for the inverse wave action cascade is a coincidence and we expect the other properties of the system to be different in this regime. Finally, when $\chi \gg 1$, we enter the wave-unaffected state ($G_\textrm{odd}=0$), for which a naive dimensional estimate yields
\begin{equation}\label{eq:spectrum_strong_elastic}
    \Pi^N \sim \frac{k N(k)}{\tau_\textrm{NL}}\sim G_\textrm{odd}^\textrm{NL} k^6 N^2(k) \sim \varepsilon_N \quad \Rightarrow \quad N(k) \sim \varepsilon_N^{1/2} (G_\textrm{odd}^\textrm{NL})^{-1/2} k^{-3}.
\end{equation}
In the next section, we will check these phenomenological estimates, as well as our theoretical predictions for the weak regime, using numerical simulations.
\\

\noindent {\bf Dissipation due to the shear elasticity.} As announced at the beginning of this section, let us finally compare the timescale of the dissipation due to the shear elasticity $\tau_{\rm diss}\sim 1/(G k^2)$ to the nonlinear timescale $\tau_\textrm{NL}$. In the weak wave turbulence regime, using the expression \eqref{log_corrected_spectrum} for the KZ spectrum of the inverse cascade, we find
\be \label{diss_ratio_weak}
\frac{\tau_{\rm NL}}{\tau_{\rm diss}} \sim \frac{G \ln^{1/5}(k\ell_d)}{G_{\rm odd}^{3/5} (G_{\rm odd}^{\rm NL})^{1/5} \varepsilon_N^{1/5}} \;.
\ee
In the critical balance, we have instead $\tau_\textrm{NL} \sim \tau_{\rm lin} \sim 1/(G_{\rm odd} k^2)$, which gives 
\be \label{diss_ratio_cb}
\frac{\tau_{\rm NL}}{\tau_{\rm diss}} \sim \frac{G}{G_{\rm odd}} \;.
\ee
In both cases, the ratio is independent of the scale (up to the logarithmic correction in the first case). In order to have an inertial range, $G$ should be very small -- more precisely we need $G\ll G_{\rm odd}$ in both cases (since $\tau_{\rm diss}$ should always be large compared to $\tau_{\rm NL}$) and $G\ll G_{\rm odd}^{3/5} (G_{\rm odd}^{\rm NL})^{1/5} \varepsilon_N^{1/5}$ in the weak regime. As we have already anticipated, since its effect is the same at all scales, the even elasticity cannot play the role of a small scale or large scale dissipation in this model, hence the additional dissipation term $\bm{\mathcal{D}}$ in equation \eqref{eom_odd_solid_true}.

\subsection{Numerical simulations}
\label{numerical_simulations_odd_solid}

\subsubsection{Numerical implementation} 

In order to test the theoretical predictions, we numerically simulate our toy model for the odd elastic solid, probing the different strong and weak states.

To enter into a driven-dissipative state analogous to fully developed turbulence, we include a stochastic forcing ($\bm{g}$ in equation \eqref{eom_odd_solid_true}) that injects energy through additive Gaussian uncorrelated noise at selected wavenumbers. Integration is done in Fourier space, employing a 4th order Runge-Kutta scheme. The non-linear term is treated using a pseudo-spectral approach with full dealiasing, while the stiff linear terms are integrated exactly through integrating factors. To ensure that dissipation is concentrated at the small scales, and to allow for the emergence of an inertial range, we replace the dissipative elastic term $G \partial^2 \bm{u}/\partial z^2$ by a hyperdissipation $\bm{\mathcal{D}}=\tilde{G}_n \partial^{2n}\bm{u}/\partial z^{2n}$ with $n=3$. To be able to dissipate also the inverse cascade, we also include a small hypodissipation, acting at large scales, by including a term $\bm{\mathcal{D}}=\tilde{G}_m^{\textrm{hypo}} \partial^{2m} \bm{u}/\partial z^{2m}$ with $m=-1$.
From these terms one can estimate respectively the small scale dissipation range $k\gtrsim k_+$ and large scale dissipation range $k\lesssim k_-$, by comparing the nonlinear and dissipation timescales as in \eqref{diss_ratio_weak} or \eqref{diss_ratio_cb} depending on the regime. For instance, in the weak wave regime we find $k_-\sim (\tilde G_m^{\rm hypo}/(G_{\rm odd}^{3/5} (G_{\rm odd}^{\rm NL})^{1/5} \varepsilon_N^{1/5}))^{1/(2-2m)}$, and in the critical balance regime $k_+\sim (G_{\rm odd}/\tilde G_n)^{1/(2n-2)}$.

In summary, the simulated PDE is 
\bea
    \frac{\partial}{\partial t} \begin{pmatrix}u_x\\u_y\end{pmatrix} 
    = 
    \begin{pmatrix}0 & G_\textrm{odd} \\ -G_\textrm{odd} & 0\end{pmatrix} \frac{\partial^2}{\partial z^2} \begin{pmatrix}u_x\\u_y\end{pmatrix}
    +
    G_\textrm{odd}^\textrm{NL} \frac{\partial}{\partial z}\left[ \left(\left[\frac{\partial u_x}{\partial z}\right]^2 + \left[\frac{\partial u_y}{\partial z}\right]^2 \right) \begin{pmatrix}0 & 1 \\ -1 & 0\end{pmatrix} \frac{\partial}{\partial z}\begin{pmatrix}u_x\\u_y\end{pmatrix} \right]
    \nn\\+
    \tilde{G}_n \frac{\partial^{2n}}{\partial z^{2n}} \begin{pmatrix}u_x\\u_y\end{pmatrix}
    +
    \tilde{G}_m^{\textrm{hypo}} \frac{\partial^{2m}}{\partial z^{2m}} \begin{pmatrix}u_x\\u_y\end{pmatrix}
    +
    \begin{pmatrix}f_x\\f_y\end{pmatrix},
\eea
where $(f_x,f_y)$ is the driving force, injecting wave action at a constant rate $\varepsilon_N$.

Without loss of generality, we can set $G_\textrm{odd}^\textrm{NL}\equiv 1$. The length of the domain is $2\pi$ and we assume periodic boundaries. We consider three distinct regimes: (I) a strong wave-unaffected regime, where the wave sustaining odd elasticity $G_\textrm{odd}$ is negligibly small w.r.t. the non-linearity, (II) a critical balance regime with intermediate $G_\textrm{odd}$, and (III) a weak wave turbulence regime where the $G_\textrm{odd}$ term dominates over the non-linear interactions. To unveil both possible inverse and forward cascades with a maximal inertial range, we consider both injection at low wavenumbers and at high wavenumbers $k_f$. Because of the high order of the wave interactions, extremely long simulation time is needed to fully develop the weak wave turbulent cascade. The total duration of the simulation is up to $T_\textrm{tot}=10^9 dt$, where $dt$ is the timestep, chosen to be smaller than the period of the fastest resolved odd wave. The spatial resolution is set to $N=128$. The full set of parameters is provided in Tab.~\ref{tab:input_odd_elastic}.

\begin{table}[h]
\begin{threeparttable}
\caption{\label{tab:input_odd_elastic}Input parameters used for the simulations of the odd elastic solid in this work.
Provided are the domain length $L$ the forcing wavenumbers $k_f$, the wave action injection rate $\varepsilon_N$, the linear odd elasticity $G_\textrm{odd}$, the non-linear odd elasticity $G_\textrm{odd}^\textrm{NL}$, the hyperdissipation $\tilde{G}_n$ with power $n$, the hypodissipation $\tilde{G}_m^{\textrm{hypo}}$ with power $m$, the integration timestep $dt$, spatial resolution $N$ and total duration $T_\textrm{tot}$.}
\begin{ruledtabular}
\begin{tabular*}{0.98\linewidth}{@{\extracolsep{\fill}}lccccccccccccc}
&&$L$&$k_f$&$\varepsilon_N$&$G_\textrm{odd}$&$G_\textrm{odd}^\textrm{NL}$&$\tilde{G}_n$&$n$&$\tilde{G}_m^{\textrm{hypo}}$&$m$&$dt$&$N$&$T_\textrm{tot}$\\
\midrule
Large-scale forcing&(I)& $2\pi$ & $[1,2]$ & 0.01 & $1.3\times10^{-3}$ & 1 & $3.2\times10^{-7}$ & 3 & 0 & N/A & $3.9\times10^{-5}$ & 128 & $5.6\times10^{2}$\\
&(II)& $2\pi$ & $[1,2]$ & 0.1 & $3.2\times10^{-1}$ & 1 & $3.2\times10^{-7}$ & 3 & 0 & N/A & $3.9\times10^{-5}$ & 128 & $5.6\times10^{2}$\\
&(III)& $2\pi$ & $[5,6]$ & 0.001 & $5.1\times10^{1}$ & 1 & $3.2\times10^{-10}$ & 3 & $2.6\times10^{-2}$ & -1 & $1.1\times10^{-5}$ & 128 & $1.2\times10^{4}$\\
Small-scale forcing&(I)& $2\pi$ & $[21,24]$ & 0.02 & $1.3\times10^{-3}$ & 1 & $3.2\times10^{-7}$ & 3 & 0 & N/A & $3.9\times10^{-5}$ & 128 & $5.6\times10^{2}$\\
&(II)& $2\pi$ & $[21,24]$ & 0.2 & $3.2\times10^{-1}$ & 1 & $3.2\times10^{-7}$ & 3 & 0 & N/A & $3.9\times10^{-5}$ & 128 & $5.6\times10^{2}$\\
&(III)& $2\pi$ & $[21,24]$ & 0.002 & $5.1\times10^{1}$ & 1 & $3.2\times10^{-10}$ & 3 & $2.6\times10^{-2}$ & -1 & $1.1\times10^{-5}$ & 128 & $1.2\times10^{4}$\\
\end{tabular*}
\end{ruledtabular}
\end{threeparttable}
\end{table}

\vspace{-0.25cm}

\begin{figure*}[ht]
    \centering
    \includegraphics[width=0.95\linewidth]{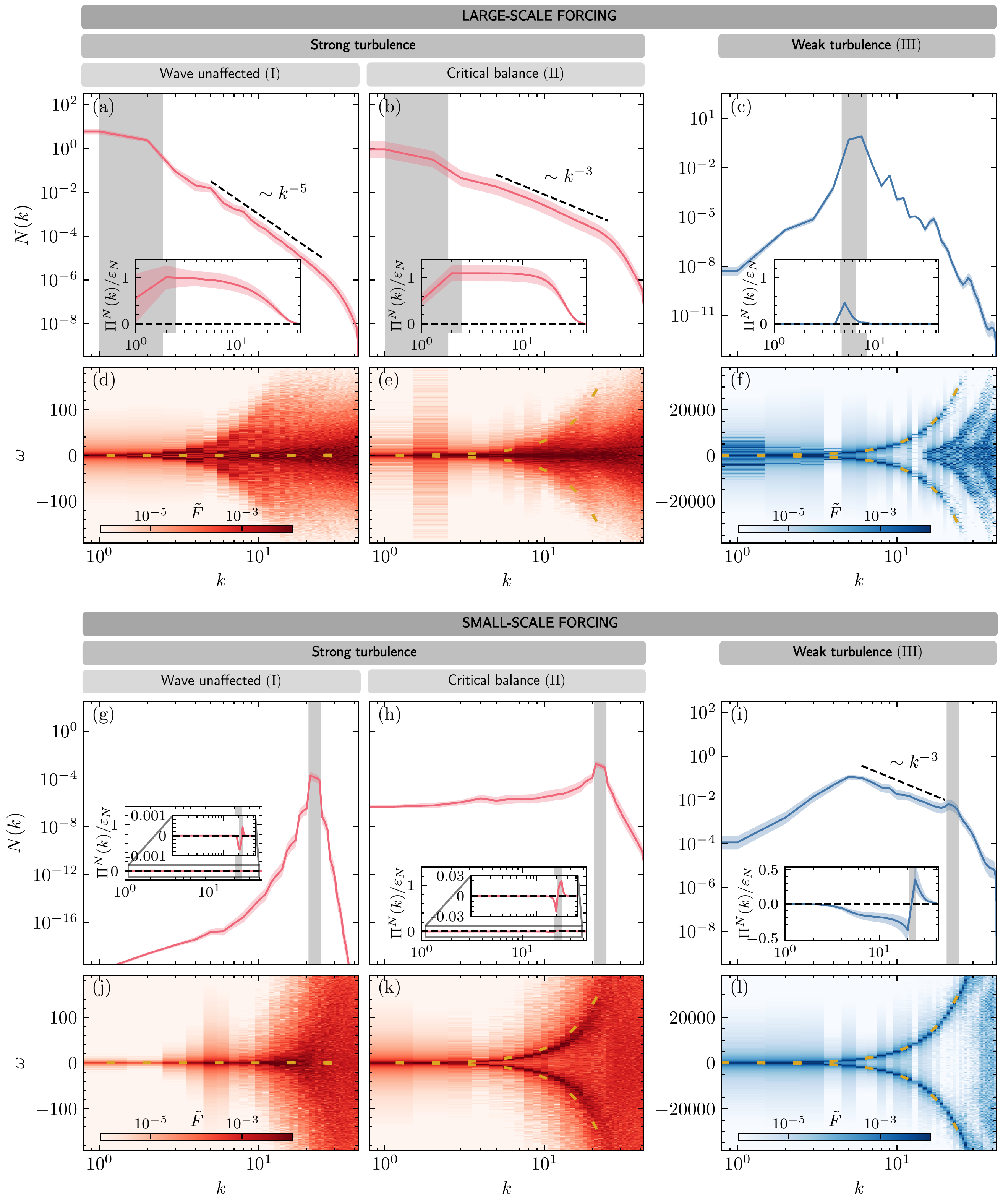}
    \caption{
    \textbf{Numerical results for the model odd elastic solid.} We consider three regimes: two regimes in a strong state, being a wave unaffected regime (first column) and a critical balance regime (second column), as well as a weak wave turbulence regime (third column). In all regimes, we consider a large-scale forcing (top half) and a small-scale forcing (bottom half). (a-c, g-i) Depict the wave action spectra $N(k)$, with insets showing the corresponding flux $\Pi^N(k)$. Gray shaded areas denote the forcing scales, while colored shaded areas depict statistical uncertainty. (d-f,j-l) provide the corresponding normalized space-time spectra of the wave action $\tilde{F}(k,\omega)$, where orange dashed lines depict the dispersion relation of odd elastic waves $\omega=\pm G_\textrm{odd} k^2$.}
    \label{fig:odd_elastic_full}
\end{figure*}

\begin{figure*}[ht]
    \centering
    \includegraphics[width=0.75\linewidth]{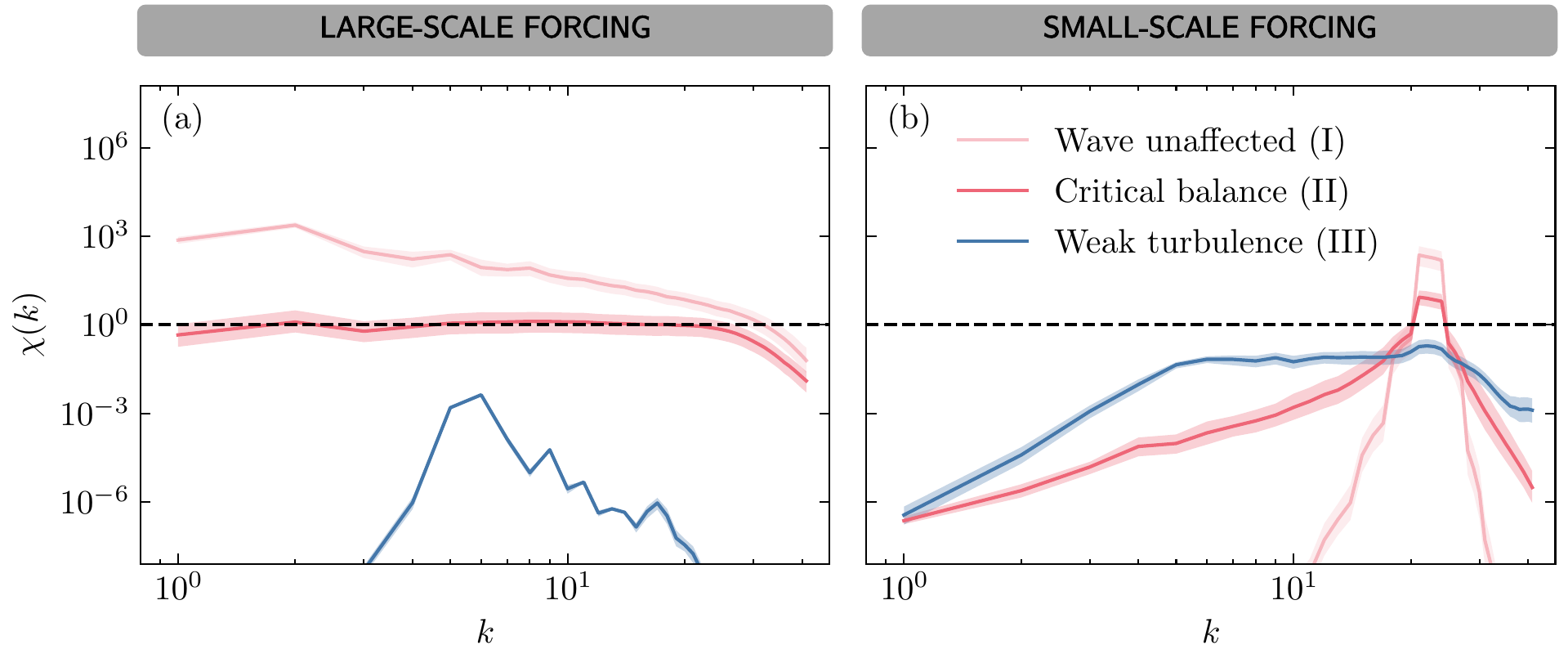}
    \caption{
    \textbf{Comparison of timescales.}
    Timescale ratio $\chi \equiv \tau_{\textrm{lin}} / \tau_{\textrm{NL}}$ comparing timescales of odd elastic waves and the non-linearity for the case of large-scale forcing (a) and small scale forcing (b) in the three considered regimes.}
    \label{fig:odd_elastic_chi}
\end{figure*}

\subsubsection{Results} 

The results are provided in Fig.~\ref{fig:odd_elastic_full}. This shows the time-averaged and spatio-temporal spectra of the wave action $N(k)$, as well as the flux $\Pi^N(k)$ thereof. Note that the flux can be obtained in the standard way from the inner product between the amplitude $\bm{u}$ and the non-linear term in Fourier space.

We find that in the strong regime, two distinct behaviors are obtained, as anticipated, which can be identified as the wave unaffected (I) and critical balance (II) regime, as supported by the measure of the timescale ratio $\chi$, see Fig~\ref{fig:odd_elastic_chi}. We find that both regimes support a forward flux of wave action and no measurable inverse flux. For the wave unaffected regime, we obtain a spectrum that empirically is close to a scaling $\sim k^{-5}$, not in agreement with the naive dimensional prediction Eq.~\eqref{eq:spectrum_strong_elastic}. The origin of this difference remains an open question, but can possibly be attributed to non-local interactions. On the other hand, for the critical balance regime, we obtain a spectral scaling consistent with $\sim k^{-3}$, in agreement with the prediction in Eq.~\eqref{eq:spectrum_cb_elastic}. Indeed, in the strong regime, no clear imprint of the dispersion relation of odd waves can be retrieved from the spatio-temporal spectra. Only for the case of small-scale forcing in Fig.~\ref{fig:odd_elastic_full}(k), a faint imprint of the dispersion relation is visible in the range $k<k_f$, which we attribute to a tiny weak wave turbulent cascade in this range (see next paragraph).

In the weak wave turbulent regime, we are able to confirm the presence of an inverse cascade, as anticipated from the weak wave turbulence theory. Indeed, when the forcing is applied at large wavenumbers, we observe an inverse flux of wave action in an inertial range between the injection and hypodissipation scales, see Fig.~\ref{fig:odd_elastic_full}(i). The scaling of the wave action spectrum is compatible with an exponent $-3$, as expected from the KZ solution, and the energy is indeed very strongly concentrated around the dispersion relation of the odd waves. When the forcing is applied at small wavenumbers, see Fig.~\ref{fig:odd_elastic_full}(c), we do not observe any significant flux of odd energy towards the small scales on the timescale of our simulations. This is compatible with the KZ solution being non-local. Our simulations do not allow us to obtain a precise estimate for the wave action spectrum in this case.

\section{Conclusion}
\label{sec:conclusion}

In this paper, we have applied weak wave turbulence theory to two parity-violating nonequilibrium models, odd fluids and solids. 
Both systems are out of equilibrium even in the absence of a forcing at a specific scale. 
When such a forcing is present, turbulent cascades across scales develop on top of a nonequilibrium background in both cases.
Despite their similarities, these two systems exhibit very different properties. In particular, while kinetic energy is conserved for the fluid, irrespective of the presence of odd viscosity, this is not the case for the solid where elastic energy is not conserved by odd elasticity. 
Nevertheless, in both cases we were able to derive the wave kinetic equation, to obtain the Kolmogorov-Zakharov spectra and to show the existence of a turbulent cascade. In the case of the fluid, we find a forward cascade of energy, while for the elastic solid it is an inverse cascade of an emergent quantity named the wave action. These results, which are all confirmed by numerical simulations, demonstrate that wave turbulence can emerge in such nonequilibrium systems. Although the turbulent cascades found in our models exhibit distinct features (such as the persistence of weak wave turbulence at all scales without a transition to strong turbulence), we have also shown that standard weak wave turbulence theory is still the right tool to study these phenomena.

\begin{acknowledgments}

This work is supported by the Netherlands Organization for Scientific Research (NWO) through the use of supercomputer facilities (Snellius) under Grant No. 2023.026. This publication is part of the project “Shaping turbulence with smart particles” with Project No. OCENW.GROOT.2019.031 of the research program Open Competitie ENW XL which is (partly) financed by the Dutch Research Council (NWO) and of the ANR project SCASCADE (grant ANR-25-CE30-5080).
SG is supported by the Simons Foundation (Grant No. 651461, PPC). 
M.F., and V.V acknowledge support from the France Chicago center through a FACCTS grant.
V.V. acknowledges support from the Army Research Office under grant nos. W911NF-22-2-0109 and W911NF-23-1-0212, from the National Science Foundation under grant no. DMR-2118415, through the Center for Living Systems (grant no. 2317138) and the National Institute for Theory and Mathematics in Biology (NITMB), from the UChicago Materials Research Science and Engineering Center (NSF DMR-2011864) and from the Theory in Biology program of the Chan Zuckerberg Initiative. 
SG acknowledges V. David for useful discussions. 
\end{acknowledgments}

\clearpage

\appendix

\section{Overdamped limit} 
\label{app:overdamped}

In this appendix we derive explicitly the overdamped limit of the equation of motion \eqref{eom_incomp_soft_solid} for the odd elastic solid. One way to do this is through the Fokker-Planck equation, as done for instance in \cite{gardiner1985}. Here we follow instead the approach used in \cite{LorentzForce}, working directly with the stochastic equation. We start by rewriting the equation \eqref{eom_incomp_soft_solid} as
\bea
    \label{eom_solid_u}
    &&\partial_t \bm{u} = \bm{p} \;, \\
    &&\rho_0 \partial_t \bm{p} = - \Gamma \bm{p} + \bm{F}(\{\partial_z^n \bm{u}\}) + \bm{f}(t,z) \;. \label{eom_solid_p}
\ea
For our model, the force $\bm{F}(\{\partial_z^n \bm{u}\})$ is defined as
\be
\bm{F}(\{\partial_z^n \bm{u}\}) = \partial_z \big[(\mu + \mu_{\text{odd}} \bm{\epsilon}) \partial_z \bm{u} + (\mu^{\text{NL}} + \mu^{\text{NL}}_{\text{odd}} \bm{\epsilon}) \lVert \partial_z \bm{u} \rVert^2 \partial_z \bm{u} \big] \;,
\ee
but in general it could be any function of $\bm{u}(t,z)$ and its spatial derivative. The driving force $\bm{f}(t,z)$ takes the form of Gaussian noise which is white in time, but correlated in space,
\be \label{corr_f}
\langle f^i(t,z) f^j(t',z') \rangle = \delta_{ij} \delta(t-t') C(z-z') \;, 
\ee
with $i,j=x,y$ and $C(z)$ an arbitrary function. Our goal is to eliminate the velocity field $\bm{p}$. For this, we integrate the equation \eqref{eom_solid_p} to obtain (we assume $\bm{p}(t=0,z)=\bm{0}$ for simplicity)
\be
\bm{p}(t,z) = \frac{1}{\rho_0} \int_0^t ds \, e^{-\frac{\Gamma}{\rho_0}(t-s)} [ \bm{F}(\{\partial_z^n \bm{u}(s,z)\}) + \bm{f}(s,z)] \;.
\ee
Inserting into the first equation \eqref{eom_solid_u}, this gives
\be \label{dtu_exact}
\partial_t \bm{u}(t,z) = \frac{1}{\rho_0} \int_0^t ds \, e^{-\frac{\Gamma}{\rho_0}(t-s)} \bm{F}(\{\partial_z^n \bm{u}(s,z)\}) + \bm{g}(t,z) \;,
\ee
where
\be \label{def_noise_g}
\bm{g}(t,z) = \frac{1}{\rho_0} \int_0^t ds \, e^{-\frac{\Gamma}{\rho_0}(t-s)} \bm{f}(s,z) \;.
\ee
Since $\bm{f}(t,z)$ is Gaussian, the new noise $\bm{g}(t,z)$ is also Gaussian.

We now take the limit where the characteristic time $\rho_0/\Gamma$ is much smaller than the timescale of the variations of $\bm{u}$, which in our case is $\omega^{-1}$. We also assume $t\gg \Gamma/\rho_0$. In this case, we can approximate the function $\bm{F}(\{\partial_z^n \bm{u}(t,z)\})$ as a constant in the integral in \eqref{dtu_exact}, leading to
\be \label{dtu_approx}
\partial_t \bm{u}(t,z) \simeq \frac{1}{\Gamma} \bm{F}(\{\partial_z^n \bm{u}(t,z)\}) + \bm{g}(t,z) \;.
\ee
We now only need to study the correlations of the Gaussian noise $\bm{g}(t,z)$. For this we use the definition \eqref{def_noise_g} and the equation \eqref{corr_f} and write
\bea
\langle g^i(t,z) g^j(t',z') \rangle &=& \frac{1}{\rho_0^2} \int_0^t ds \int_0^{t'} ds' e^{-\frac{\Gamma}{\rho_0}(t+t'-s-s')} \langle f^i(t,z) f^j(t',z') \rangle \nn \\
&=& \frac{1}{\rho_0^2} \delta_{ij} C(z-z') \int_0^t ds \int_0^{t'} ds' e^{-\frac{\Gamma}{\rho_0}(t+t'-s-s')} \delta(s-s') \nn \\
&=& \frac{1}{\rho_0^2} \delta_{ij} C(z-z') \int_0^{\min(t,t')} ds e^{-\frac{\Gamma}{\rho_0}(t+t'-2s)} \nn \\
&\simeq& \frac{1}{2\rho_0\Gamma} \delta_{ij} C(z-z') e^{-\frac{\Gamma}{\rho_0}|t-t'|} \;.
\eea
This noise is correlated in time over a timescale $\rho_0/\Gamma$. However, if this timescale is much smaller than any relevant timescale in the system 
(which in our case again amounts to $\rho_0/\Gamma\ll\omega^{-1}$), we can approximate the exponential as a delta function and write
\be \label{corr_g}
\langle g^i(t,z) g^j(t',z') \rangle = \frac{1}{\Gamma^2} \delta_{ij} \delta(t-t') C(z-z') \;.
\ee
Thus, we simply have $\bm{g}(t,z)=\bm{f}(t,z)/\Gamma$, which concludes the derivation of the overdamped equation \eqref{eom_incomp_soft_solid_overdamped}.

\bibliography{pre}

\end{document}